\documentclass[twocolumn,superscriptaddress,amsmath,amssymb,showpacs,prl]{revtex4-2}

\usepackage{graphicx}
\usepackage{color}
\usepackage{makecell}
\usepackage{footnote}
\renewcommand{\phi}{\varphi}
\newcommand{\myeqref}[1]{Eq.\ref{#1}}

\begin{document}

\title{Intermediate spin state and the B1-B2 transition in ferropericlase at tera-Pascal pressures}

\author{Tianqi Wan}
	\affiliation{Department of Applied Physics and Applied Mathematics, Columbia University, New York, NY 10027, USA}

\author{Yang Sun}
	\affiliation{Department of Applied Physics and Applied Mathematics, Columbia University, New York, NY 10027, USA}

\author{Renata M. Wentzcovitch}
	\email{rmw2150@columbia.edu}
	\affiliation{Department of Applied Physics and Applied Mathematics, Columbia University, New York, NY 10027, USA}
	\affiliation{Department of Earth and Environmental Sciences, Columbia University, New York, NY 10027, USA}
	\affiliation{Lamont–Doherty Earth Observatory, Columbia University, Palisades, NY 10964, USA}

\date{Sep. 21, 2021}

\begin{abstract}

Ferropericlase (fp), (Mg$_{1-x}$Fe$_x$)O, the second most abundant mineral in the Earth's lower mantle, is expected to be an essential component of super-Earths' mantles. Here we present an ab initio investigation of the structure and magnetic ground state of fp up to $\sim$ 3 TPa with iron concentrations (x$_{Fe}$) varying from 0.03 to 0.12. Calculations were performed using LDA+U$_{sc}$ and PBE exchange-correlation functionals to elucidate the pressure range for which the Hubbard U (U$_{sc}$) is required. Similar to the end-members FeO and MgO, fp also undergoes a B1 to B2 phase transition that should be essential for modeling the structure and dynamics of super-Earths' mantles. This structural transition involves a simultaneous change in magnetic state from a low spin (LS) B1 phase with iron total spin S=0 to an intermediate spin (IS) B2 phase with S=1. This is a rare form of pressure/strain-induced magnetism produced by local cation coordination changes. Phonon calculations confirm the dynamical stability of the iron B2-IS state. Free energy calculations are then carried out including vibrational effects and electronic and magnetic entropy contributions. The phase diagram is then obtained for low concentration fp using a quasi-ideal solid solution model. For x$_{Fe}$ $>$ 0.12 this approach is no longer valid. At ultra-high pressures, there is an IS to LS spin state change in Fe in the B2 phase, but the transition pressure depends sensitively on thermal electronic excitations and on x$_{Fe}$. 

\end{abstract}

\maketitle

\section{I. Introduction}
Ferropericlase (fp), i.e., rocksalt-type (Mg$_{1-x}$Fe$_{x}$)O (x$_{Fe} < 0.5 $ ) is a major phase of the Earth's lower mantle defined by the 660 km discontinuity produced by the post-spinel transition \cite{1}. Based on the expected composition of rocky exoplanets \cite{2}, it is also expected to exist in the mantles of Super-Earths \cite{3,4}, i.e., planets with up to $\sim$ 14 Earth masses (M$_\oplus$). These planets are among the most relevant in searching for habitable planets and extra-terrestrial life; therefore, there is great interest in understanding their internal structure and geophysical properties. Considerable effort is dedicated to discovering their constituent phases and their properties. Such phases are primarily novel silicates, oxides, and iron alloys involving Mg, Al, Ca, Fe, Si, H, S, and C. The first step has been to assume compositions close to that of the Earth and other terrestrial solar planets \cite{2} and add complexity as knowledge advances.

The properties of fp are well studied within the Earth pressure range. For instance, iron in fp undergoes a spin-state change from a high spin (HS) with total spin S=2 to a low spin (LS) with S=0 \cite{5,6}. However, on Earth, pressure and temperature at the core-mantle boundary (CMB) are only 0.135 TPa and ~ 4,000 K. In some terrestrial exoplanets, CMB conditions might reach several TPa and $\sim$ 10,000 K \cite{4}. Characterization of fp on a wider range of pressure and temperature conditions is essential to better understand terrestrial mantles as well as the behavior of iron in related phases at similar conditions.

On Earth’s mantle, fp exists in the NaCl-type (B1) structure \cite{7}, but at approximately 600 GPa, MgO undergoes a transition from the B1 (NaCl-type) to the B2 (CsCl-type)\cite{8,9}. Therefore, it is natural to expect the B2 phase to occur in fp at ultrahigh pressures, as recently proposed \cite{10}. The complexity of the structural change coupled to a possible spin-state change in fp makes the B1 structure’s stability field largely unknown and challenging to pin down at ultra-high temperatures and pressures.

The presence of localized 3\textit{d} electrons in iron requires methods that go beyond standard density functional theory (DFT) to address their strongly correlated nature \cite{11,12,13,14,15,16}. DFT+U is a popular method that adds the Hubbard correction to standard DFT calculations \cite{16,17}. The reliability of DFT+U results depends greatly on the Hubbard parameter U. To be predictive, U should be determined by ab initio \cite{12}, self-consistently \cite{18}, and be structure- and spin-state dependent \cite{19,20,21,22,23,24}. Moreover, electron delocalization under pressure induces an insulator to metal transition (IMT) in the FeO end member (Greenberg et al., 2020, under review) and should also occur in fp at some high pressure. Therefore, the performance of DFT+U and other functionals needs to be examined carefully and consistently in both metallic and insulating states of fp.

We study the B1-B2 structural transition and iron spin states in fp up to 3 TPa using LDA+U$_{SC}$ and conventional DFT methods. The dependence of U on pressure, volume, structure, and spin-state are carefully considered in the (Mg$_{1-x}$Fe$_{x}$)O system with x$_{Fe}$ up to 0.125 at ultra-high pressures and temperatures. Various entropic contributions have also been included in the free energy and phase diagram calculations.

This paper is organized as follows. In Section II we present the details of computational methods. Section III offers an analysis of the electronic structure, including spin states, free energy calculations, and phase diagrams. Section IV presents our conclusions.

\section{II. Methods}
\subsection{2.1 \textit{Ab initio} calculations}
\textit{Ab initio} calculations are carried out with the Quantum ESPRESSO code\cite{26,27}. Local density approximation (LDA) and LDA+U$_{SC}$ calculations use Vanderbilt's ultra-soft pseudopotentials \cite{28}  with valence electronic configurations 3s$^2$3p$^6$3d$^{6.5}$4s$^1$4p$^0$ and 2s$^2$2p$^4$ for Fe and O, respectively. The pseudopotential for Mg was generated by von Barth-Car’s method using five configurations 3s$^2$3p$^0$, 3s$^1$3p$^1$, 3s$^1$3p$^{0.5}$3d$^{0.5}$, 3s$^1$3p$^{0.5}$, and 3s$^1$3d$^1$ with decreasing weights 1.5, 0.6 0.3 0.3, and 0.2, respectively. These pseudopotentials for Fe, Mg, and O were generated, tested, and previously used \cite{29}.  Perdew-Burke-Ernzerhof (PBE) \cite{30} calculations are carried out using the projector augmented wave (PAW) method with PAW dataset from the pslibrary \cite{31}. Plane-wave energy cutoffs are 100 Ry and 1,000 Ry for electronic wave functions and spin-charge density and potentials, respectively. The irreducible Brillouin zone of the 64-atom supercell is sampled by a 6×6×6 Monkhorst-Pack mesh \cite{32} when computing the charge density. Effects of larger energy cutoff and k-point sampling on calculated properties are insignificant. The convergence thresholds are 0.01 eV/\AA\ for all components of all forces(f$_x$, f$_y$, and f$_z$), including atomic forces and averaged forces in the supercell, and 1×10$^{-6}$ eV for the total energy of the supercell, 64 atoms in this work. The Mermin functional is used in all calculations \cite{33,34} to address the important effect of thermal electronic excitation on the free energy and dynamical stability of the phases. 

\begin{figure}
\includegraphics[width=0.5\textwidth]{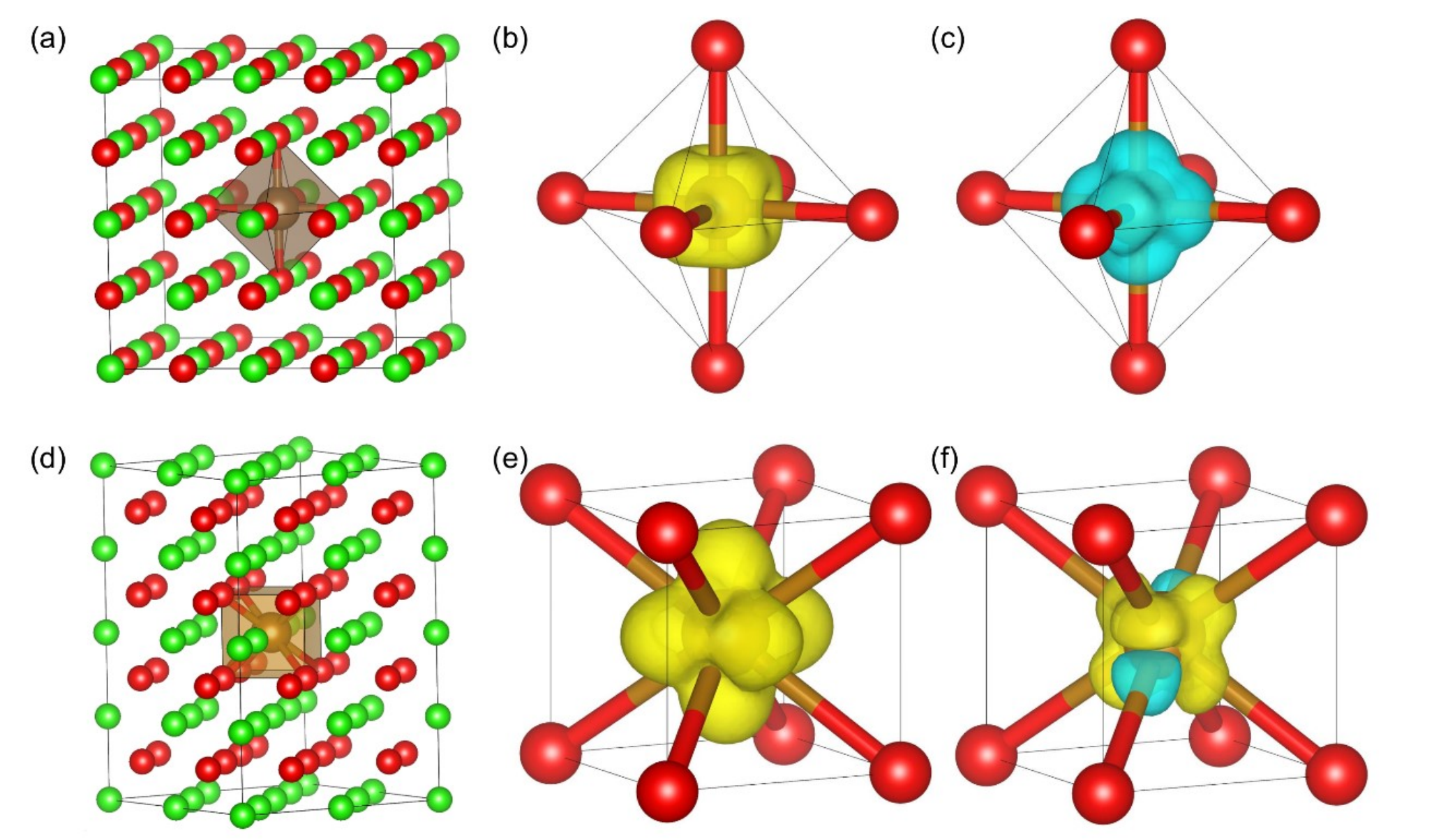}
\caption{\label{fig:fig1} Atomic structures of (a) B1 and (d) B2 phases with x$_{Fe}$= 0.03125; (b) fully occupied t$_{2g}$ orbitals and (c) unoccupied e$_g$ orbitals of LS Fe$^{2+}$ in the octahedral crystal field of the B1 structure; (e) fully occupied e$_g$ orbitals and (f) half occupied e$_g^{'}$ (formerly t$_{2g}$) of IS Fe$^{2+}$ in the cubic crystal field; yellow and blue represent occupied, fully or partially, and unoccupied orbitals,respectively. Green, brown, and red represent Mg, Fe, and O, respectively.}
\end{figure}

In LDA+U$_{SC}$ calculations, the Hubbard correction \cite{16} is applied to Fe-3\textit{d} states. The Hubbard parameter U is computed using density-functional perturbation theory (DFPT) \cite{35}. The convergence threshold for the response function is 1×10$^{-6}$  eV. We employed an automated iterative scheme to obtain the self-consistent U$_{SC}$ parameter while simultaneously optimizing the structure and spin state \cite{22}. Starting from an empirical U = 4.3 eV, we compute energies corresponding to all possible occupation matrices for a particular spin state. The electronic configuration, i.e., occupation matrix with the lowest energy, is then selected to further structural optimization using variable cell shape molecular dynamics \cite{36,37}. Then a new U parameter is recalculated for further structural optimization. The process continues until mutual convergence of structure and U is achieved \cite{22} for a convergence threshold of 0.01 eV for the U parameter and the convergence criteria mentioned above for structural optimizations. For the metallic states, we use the Mermin functional in calculations a \textit{posteriori}, i.e., without further changing U from self-consistent calculations.
Phonon calculations are performed in 64-atom supercells using the finite-displacement method with the Phonopy code \cite{38} and LDA+U$_{SC}$ forces obtained with the Quantum ESPRESSO. All phonon calculations used the Mermin functional \cite{33,34,39}. An unshifted  2×2×2  q-mesh was used to compute phonon frequencies using DFPT in a 64-atom cell. Vibrational density of states (VDoSs) are then obtained using a denser q-point 6×6×6 mesh. The vibrational contribution to the free energy is calculated using the quasi-harmonic approximation \cite{40}.

\subsection{2.2 Quasi-ideal solid solution model}

We address the B1-B2 phase boundary using a binary solid solution model without computing the free energy of the FeO end-member. The MgO-FeO system does not form an ideal solid solution (ISS). FeO-MgO interactions can not be neglected, and we do not do so. Supercell calculations naturally include the MgO-FeO interaction energy, even using a single atomic configuration. For x$_{Fe}$ $<$ 0.125, iron atoms are well separated on average, and the energy difference between various supercell atomic configurations can be disregarded with out significant effects. Fe atoms are surrounded by O, and Mg is most likely the second nearest neighbor(see structures in Fig.S5 in the Supplemental Material). Given this situation, a single atomic configuration of fp with small x$_{Fe}$ can be considered as an ed member of a dilute solid solution. Therefore, instead of using the free energy of the FeO and MgO in the ISS modeling, one can calculate the free energy for some x$_{Fe}$(x$_{Fe}$ $<$ 0.125 in this paper) using MgO and (Mg$_{0.875}$Fe$_{0.125}$)O as end members.

Specifically, in an A-B binary system (MgO-FeO in the current case) at a given pressure and temperature, the Gibbs free energy of mixing curves of B1 and B2 ISSs are

\begin{widetext}

\begin{equation}
G_{x_B}^{B1}=k_BT[x_B\ln x_B+(1-x_B)\ln (1-x_B)]+x_B(\Delta G_{B}^{B1-B2}), \label{1a}
\end{equation}

\begin{equation}
G_{x_B}^{B1}=k_BT[x_B\ln x_B+(1-x_B)\ln (1-x_B)]-(1-x_B)(\Delta G_{A}^{B1-B2}) , \label{1b}
\end{equation}

\end{widetext}

where k$_B$ as the Boltzmann constant and x$_{B}$=x$_{Fe}$/0.125, where x$_{Fe}$ is the iron concentration. Here, $\Delta G_{A}^{B1-B2}$ and $\Delta G_{B}^{B1-B2}$ are the free energy differences of the end members in the two phases involved, i.e., $\Delta G_{A}^{B1-B2}=G_{A}^{B1}-G_{A}^{B2}$ and $\Delta G_{B}^{B1-B2}=G_{B}^{B1}-G_{B}^{B2}$. End member A corresponds to MgO, while end member B is (Mg$_{0.875}$Fe$_{0.125}$)O. In this case, the ISS-like modeling includes interaction between the standard MgO-FeO end members. We refer to this level of modeling as a quasi-ISS(QISS; see Supplemental Material)

In a binary solid solution model, the compositions of the B1-B2 solvus lines are,

\begin{widetext}

\begin{equation}
\begin{aligned}
& X_B^{B2} = \frac{1-\exp{(-\frac{\Delta G_{A}^{B1-B2}}{k_BT})}}{\exp{(-\frac{\Delta G_{B}^{B1-B2}}{k_BT})}-\exp{(-\frac{\Delta G_{A}^{B1-B2}}{k_BT})}} , 
& X_B^{B1} = X_B^{B2}\exp{(-\frac{\Delta G_{A}^{B1-B2}}{k_BT})}.
\end{aligned}
 \label{1c1}
\end{equation}

\end{widetext}

The B2 mol fraction, $n_{B2}$, is given by the lever rule,

\begin{equation}
n_{B2}= \frac{x_B-X_B^{B1}}{X_B^{B2}-X_B^{B1}} . \label{1d}
\end{equation}

\section{III. Results}
\subsection{3.1 Intermediate spin state in the B2 structure}
We first investigate the electronic structure, including the spin states, of FeMg$_{31}$O$_{32}$ (x$_{Fe}$=0.03125, noted as fp3 hereafter), in the B1 and B2 phases. To create a similar 64-atom cell in the B2 phase, we constructed a $2\sqrt{2}$×$2\sqrt{2}$×4 supercell. The crystal structures are shown in Fig.~\ref{fig:fig1}(a) and (b).

According to several previous studies, the HS-LS static transition in the B1 phase occurs at $\sim$ 60 GPa in fp3 \cite{42}. Therefore, in the pressure range considered here, i.e., 300 GPa to 2 TPa, B1 fp is in the LS state. However, the spin state of fp in the B2 structure is mainly unknown. Due to different cation coordination, the 3\textit{d} energy level degeneracies in Fe$^{2+}$ in the B2 and B1 structures differ. Fe$^{2+}$ is octahedrally coordinated in the B1 phase, and the 5-fold \textit{d} levels split mainly into a lower triplet, the t$_{2g}$ states, and an upper doublet, the e$_g$ states with further smaller Jahn-Teller splitting. The e$_g$ orbitals point toward negatively charged n.n. oxygens, while the t$_{2g}$ orbitals point away from them (see Fig.~\ref{fig:fig1}(b) and (c)). Electrostatic interaction contributes to lowering the energy of t$_{2g}$ orbitals relative to those of e$_g$ orbitals in the octahedral environment. In the B2 phase, Fe$^{2+}$ has cubic coordination. The t$_{2g}$ orbitals point to the n.n. oxygens, while the e$_g$ orbitals point to the interstitial sites, away from them (see Fig.~\ref{fig:fig1}(e) and (f)). Therefore, the doublet e$_g$ becomes energetically favorable relative to the t$_{2g}$ triplet in the B2 structure. Jahn-Teller splitting is also observed in this case.

The change in the electronic energy levels in going from the B1 to the B2 structure induces changes in the spin state and Fe-O bond lengths, i.e., Jahn-Teller distortions. In the B1 phase, Fe$^{2+}$ exists in the HS and LS states with configurations \textit{d}$_{\uparrow}^5$\textit{d}$_{\downarrow}^1$ and \textit{d}$_{\uparrow}^3$\textit{d}$_{\downarrow}^3$, respectively. Two intermediate spin (IS) states, \textit{d}$_{\uparrow}^4$\textit{d}$_{\downarrow}^2$, with different occupation
matrices were shown to be metastable in Fe$^{2+}$ in B1-type fp \cite{42}. In the B2 phase, the IS state can be stabilized as a consequence of the lower e$_g$ doublet and an additional Jahn-Teller splitting in the t$_{2g}$ triplet. In this state, four electrons occupy all e$_g$ orbitals, and the remaining two electrons occupy two of the t$_{2g}$ orbitals, now e$_g^{'}$, producing a \textit{d}$_{\uparrow}^4$\textit{d}$_{\downarrow}^2$ electronic configuration with S=1. Therefore, the cubic crystal field favors the formation of the IS state. Alternatively, the two electrons in the e$_g^{'}$ manifold can occupy a single e$_g^{'}$ orbital, with spin up and down, producing the LS state with \textit{d}$_{\uparrow}^3$\textit{d}$_{\downarrow}^3$. As shown later, the B2-LS state is metastable in the cubic crystal field at lower pressures.

\begin{figure}[t]
\includegraphics[width=0.5\textwidth]{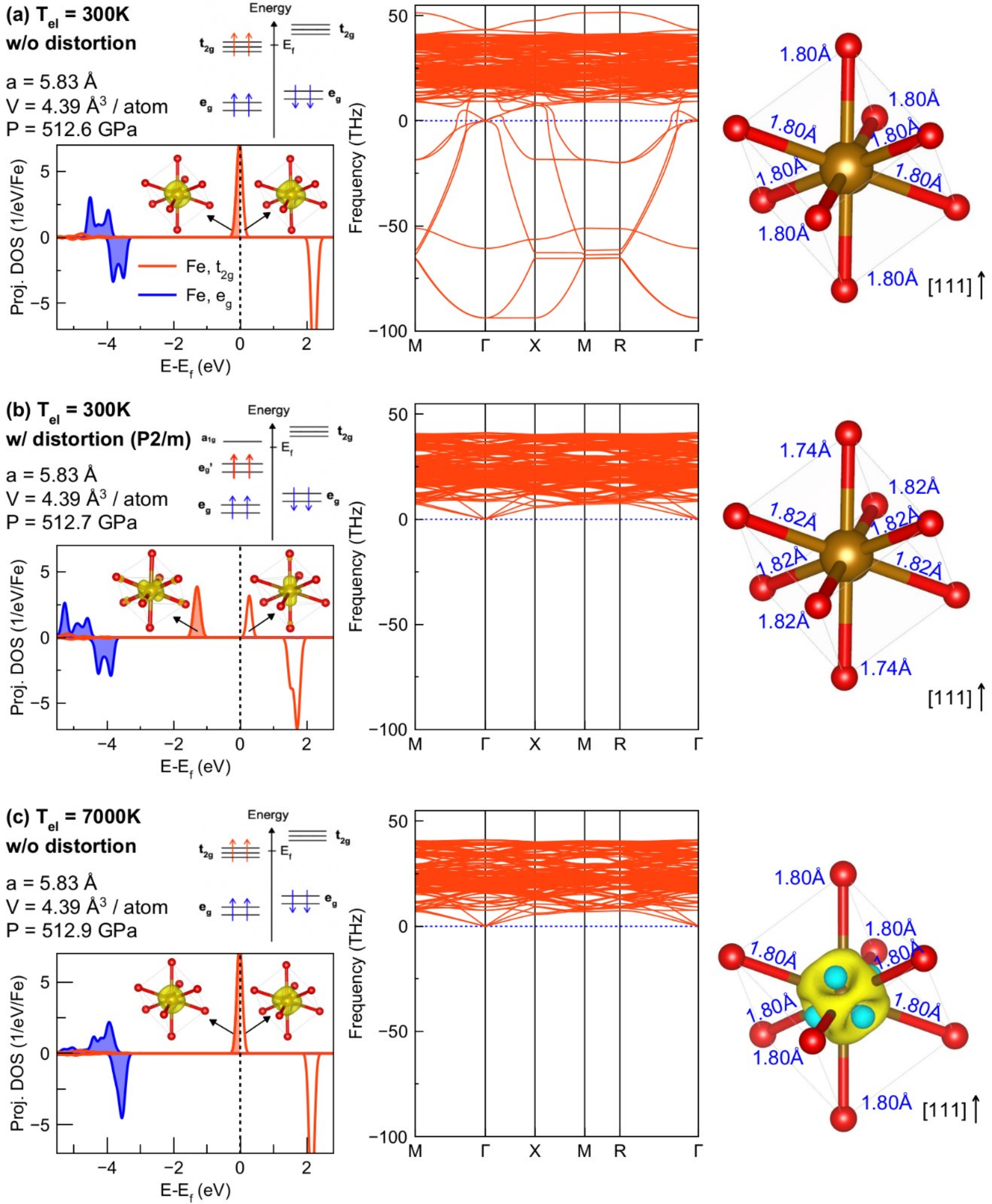}
\caption{\label{fig:fig2} Left panel: t$_{2g}$ and e$_g$ projected density of states (PDoS), charge density, and electron occupation diagram for the 3\textit{d} electrons in Fe$^{2+}$ in the B2-IS state for x$_{Fe}$=0.03125 obtained with LDA+U$_{SC}$. Middle: Corresponding phonon dispersions. Right panel:
Fe-O bond lengths and charge difference between (a) and (c). $(\Delta\rho=\rho(7,000 K)-\rho(300 K)$. The charge moves from blue to yellow regions.}
\end{figure}

Fig.~\ref{fig:fig2} shows t$_{2g}$ and e$_g$ projected density of states (PDoS) in the B2-IS phase (\textit{d}$_{\uparrow}^4$\textit{d}$_{\downarrow}^2$). Fig.~\ref{fig:fig2}(a) shows that the 3\textit{d} electron PDoS of the clamped ion structure is metallic with an electronic (Mermin) temperature T$_{el}$ = 300 K. The three t$_{2g}$-up orbitals at the Fermi level are partially occupied with $\sim$2/3 occupancy. The FeO bond lengths in this state with cubic coordination are the same. In this situation, fp3 shows imaginary phonon frequencies pointing to structural instability. We add the displacement mode of the largest imaginary frequency phonon to the undistorted structure and relax the atomic positions with T$_{el}$ = 300 K. Fig.~\ref{fig:fig2}(b) shows the resulting structure with shorter Fe-O bond lengths along [111], i.e., a local rhombohedral distortion. The t$_{2g}$ triplet splits into an a$_{1g}$ singlet and an e$_g^{'}$ doublet with this distortion. With two electrons occupying the doublet, the system becomes insulating with a bandgap of 1 eV shown in Fig.~\ref{fig:fig2}(b). In contrast, with T$_{el}$ = 7,000 K, a more realistic temperature for exoplanetary interiors, the system is metallic again, and phonon instabilities are no longer present in the cubic environment (see Fig.~\ref{fig:fig2}(c)). The right panel in Fig.~\ref{fig:fig2}(c) shows the charge distribution difference between the distorted structure (Fig.~\ref{fig:fig2}(b)) with T$_{el}$ = 300 K and the cubic structures with T$_{el}$ =7,000 K. We see a charge transfer from interstitial regions to the bonding region, which helps to delocalize the \textit{d}-electrons among the three t$_{2g}$ states.Despite the finite carrier concentration at the Fermi level, a disordered material with such a narrow electronic bandwidth may not behave as a conductor.

We note that the T$_{el}$-dependent phonon instability revealed in Fig.~\ref{fig:fig2} does not depend on the functional or iron composition. Similar results are produced with the PBE functional and with x$_{Fe}$= 0.125 (see Fig.~\ref{fig:figS1} and \ref{fig:figS2} in the Supplementary Information section). In all cases, the increasing T$_{el}$ leads to a systematic increase of phonon stability (see also Fig.~\ref{fig:figS3}-\ref{fig:figS4}). However, the uncertainty in the LDA+U$_{SC}$ gap, and electron-phonon and phonon-phonon interactions not addressed in this calculation, naturally introduce uncertainty in the T$_{el}$ necessary to remove the Jahn-Teller instability. In addition, fp must be a disordered solid solution at high temperatures; therefore, coherent rhombohedra distortions at low temperatures, as shown in Fig.~\ref{fig:fig2}(b), are less likely to form when Fe is randomly distributed. 

\subsection{3.2 B1-B2 phase transitions in fp3}
\begin{figure}
\includegraphics[width=0.48\textwidth]{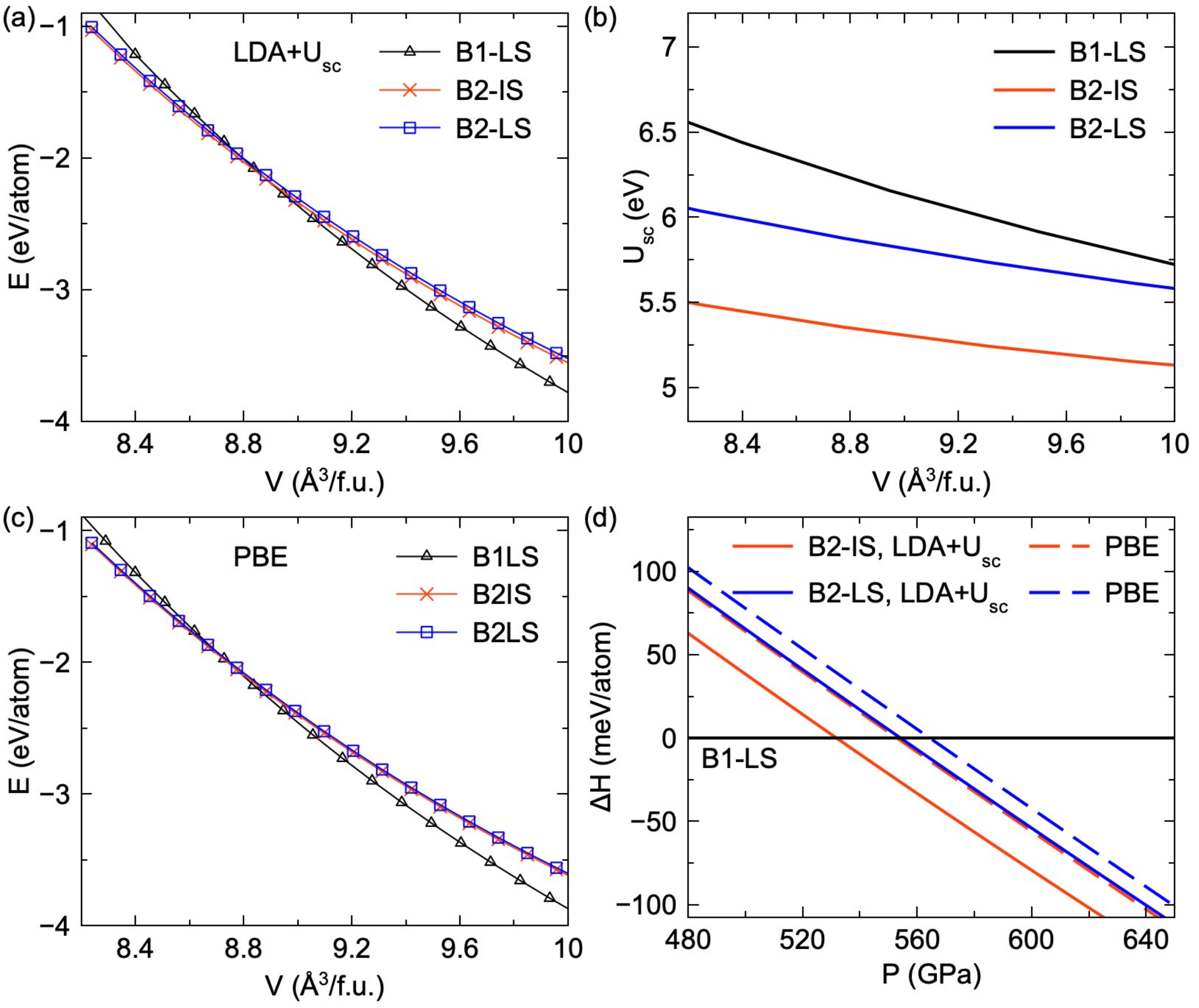}
\caption{\label{fig:fig3} Energy-volume curves for B2-IS, B2-LS and B1-LS states with x$_{Fe}$= 0.03125 in (a) LDA+U$_{SC}$ calculations and (c) PBE calculations. (b) The self-consistent Hubbard parameters U. (d) Relative enthalpies using B1-LS as the reference.}
\end{figure}

To determine the relative stability of B1 and B2 phases in fp3, we perform LDA+U$_{SC}$ and PBE calculations in a large volume (pressure) range with T$_{el}$ =1,000 K. All atomic positions are fully optimized to accommodate favorable Jahn-Teller distortions. In the relaxed structures at $\sim$ 600 GPa, we find the average Fe-O bond lengths are longer than the Mg-O ones by 0.8$\%$ in the B2-IS state and 0.9$\%$ in the B2-LS state. Fig.~\ref{fig:fig3}(a) shows the volume-dependent total energy of different states. Upon decreasing volume, the B1-LS state becomes less favored than the B2 states. Fig.~\ref{fig:fig3}(b) shows the self-consistent and structurally consistent Hubbard parameters obtained in these calculations. The B2-IS state shows systematically lower U values than the LS states. This trend is similar to that seen in the U parameters in FeO where the LS state of Fe$^{2+}$ always shows the largest self-consistent U value, regardless of crystal structure \cite{22}. The U values of B1-LS and B2-LS differ by $\sim$ 1 eV. This shows the Hubbard parameter has a substantial volume and structure dependence besides the electronic configuration dependence.  Fig.~\ref{fig:fig3}(c) shows similar PBE results indicating a very similar B1-B2 transition behavior, with PBE transition pressures being higher by 10-30 GPa than LDA+U$_{SC}$ transition pressures, a typical difference in performance between LDA and PBE functionals \cite{43}.

The energy vs. volume results of each state is fit with the Birch-Murnaghan (BM) equation of state (EoS). Detailed EoS parameters are shown in Supplementary Tables S1 and S2. Enthalpies obtained from the static BM-EoS are shown in Fig.~\ref{fig:fig3}(d). Regardless of the functional, the B2-IS phase always shows lower enthalpy than the B2-LS in the pressure range considered here, indicating that the IS state is the ground state in the B2-type phase for small x$_{Fe}$. The static B1-LS $\rightarrow$ B2-IS transition pressure is 531 GPa in LDA+U$_{SC}$ calculations and 554 GPa in PBE calculations.

\subsection{3.3 B1-B2 phase boundary up to x$_{Fe}$= 0.125}
\begin{figure}[t]
\includegraphics[width=0.45\textwidth]{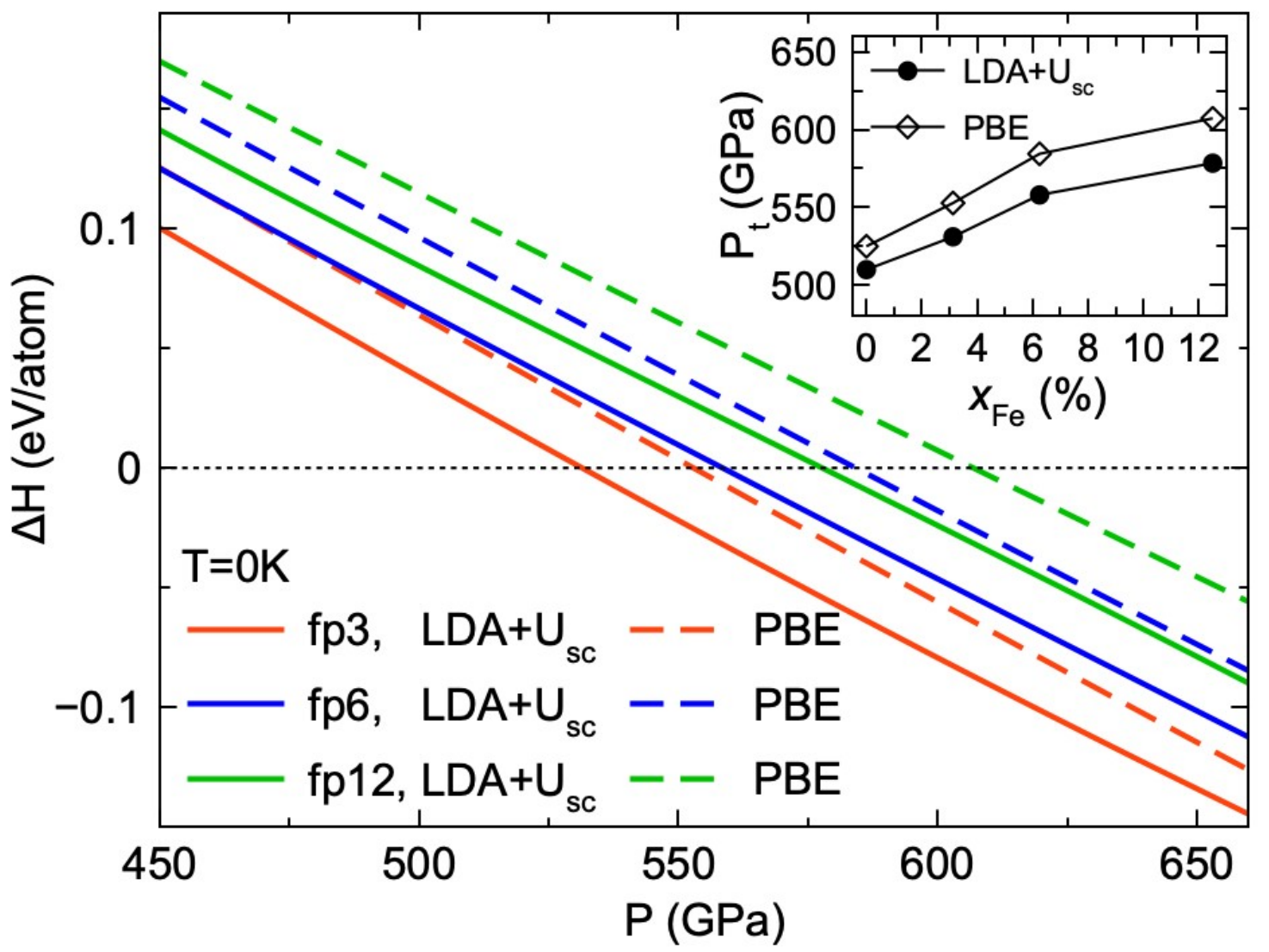}
\caption{\label{fig:fig4} Relative enthalpies between B1-LS to B2-IS states in fp3, fp6, and fp12. The black-dot line corresponds to the B1-LS state.}
\end{figure}

We further investigate the B1-LS $\rightarrow$ B2-IS transition in fp with larger iron concentrations in Fe$_2$Mg$_{30}$O$_{32}$ and Fe$_4$Mg$_{28}$O$_{32}$ (x$_{Fe}$= 0.0625 and 0.125, noted as fp6 and fp12 hereafter). It has been shown that the iron arrangement in the fp solid solution has a minor effect on the spin state change pressure in the B1 phase, at least for small x$_{Fe}$ as investigated here \cite{44}. Therefore, we distribute Fe uniformly in both B1 and B2 lattices and disregard Fe-Fe interaction effects in first-order (see Supplementary Fig.~\ref{fig:figS5}). The relative enthalpies for different iron concentrations are shown in Fig.~\ref{fig:fig4}. With increasing iron concentration, the static B1-LS $\rightarrow$ B2-IS transition pressure increases from MgO to (Fe$_{0.125}$Mg$_{0.875}$)O. PBE calculations show a systematically higher transition pressure than the LDA+U$_{SC}$ calculations, with a difference of 20-30 GPa. This difference is typical of what one expects from LDA and PBE calculations \cite{43}.

\subsection{3.4 Thermodynamic phase boundary}

We now compute the B1-LS $\rightarrow$ B2-IS thermodynamic phase boundary. We use the quasi-harmonic approximation (QHA) to compute the vibrational free energy. Supplementary Fig.~\ref{fig:figS3} and Fig.~\ref{fig:figS4} show phonon dispersions of the B2-IS phase using different functionals. It demonstrates that thermal electronic excitation with a high T$_{el}$ is required to stabilize phonons in LDA+U$_{SC}$ and PBE calculations. This stabilization is related to the stabilization of the metallic state. Electron-phonon interaction in the vibrating lattice might also decrease the gap and the required T$_{el}$ for metallization. Phonon-phonon interaction might also help to stabilize the unstable modes. A dynamical mean-field theory (DMFT) calculation may be necessary to accurately address the effect of electron-electron interaction on the metallic state\cite{25} and anharmonic phonon calculations should also elucidate these instabilities in the phonon spectrum.

\begin{figure}
\includegraphics[width=0.48\textwidth]{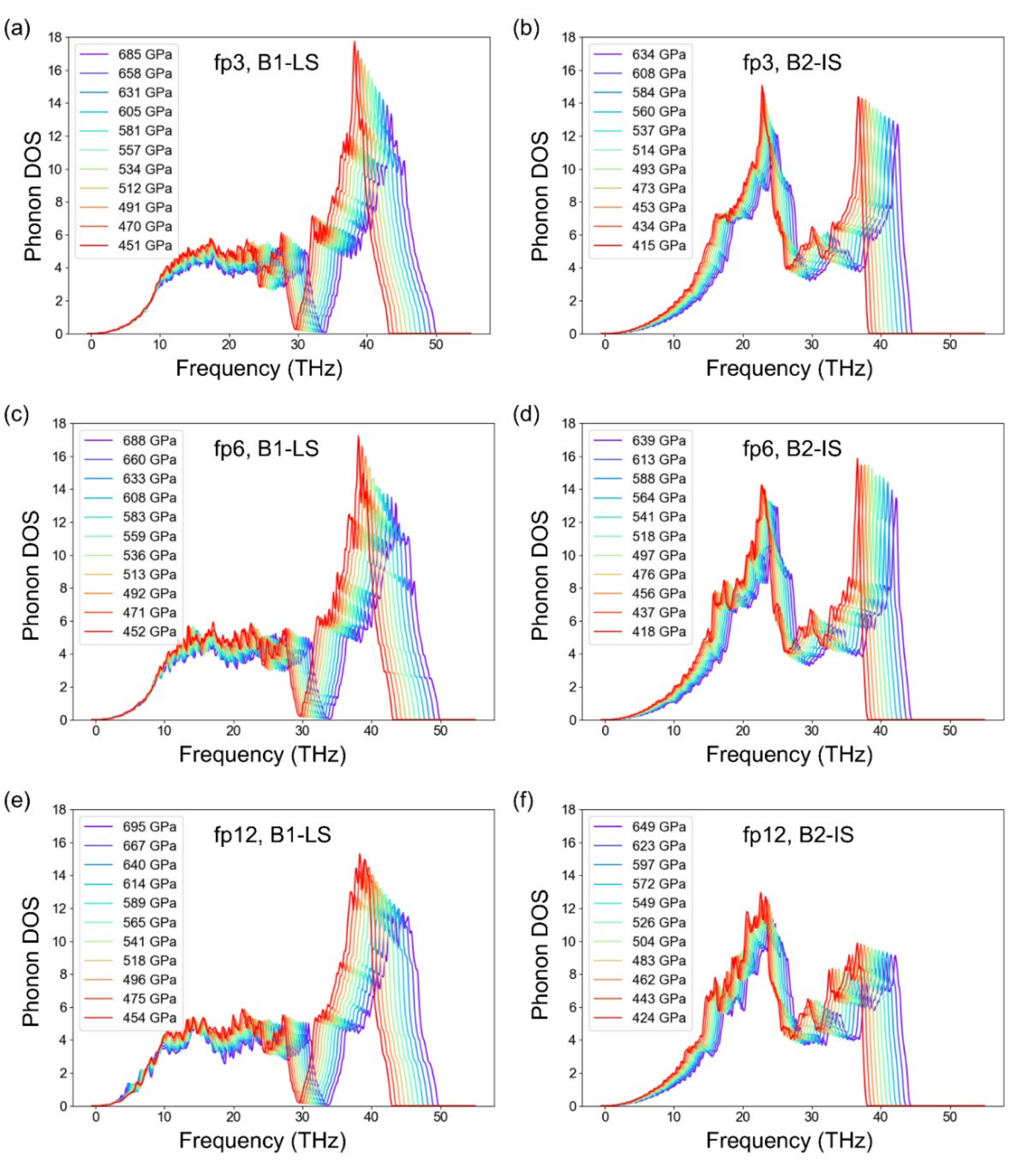}
\caption{\label{fig:fig5} Phonon density of state for (a) B1-LS of fp3; (b) B2-IS of fp3; (c) B1-LS for fp6; (d) B2-IS of fp6; (e) B1-LS of fp12; (f) B2-IS of fp12.}
\end{figure}

Nevertheless, unavoidable thermal electronic excitations in LDA+U$_{SC}$ calculations stabilize the metallic state and the phonon spectrum. As long as the phonon dispersion has no imaginary mode frequencies, the electronic temperature does not affect significantly the free energy, thermodynamic properties, and phase boundary. We have tested the effect of electronic temperature on the phase boundary and found no significant difference between Tel = 5000 K and Tel = 7000 K, as shown in Supplementary Fig. S8. Therefore, we chose to use a consistent Tel = 7000 K for all the phonon calculations shown in Fig.~\ref{fig:fig5}. This is an \textit{ad hoc} choice but an approximation that enables a first glimpse of the stability field of these phases. Furthermore, with increasing pressure, phonon frequencies increase in both B1 and B2 phases. No imaginary frequency exists at these conditions. Therefore, at high temperatures, both B2-IS and B1-LS states of fp with x$_{Fe}$ \textless \, 0.125 are dynamically stable in this calculation. 

As recently pointed out \cite{39}, calculations of thermodynamic properties for metallic systems need to fully include thermal electronic excitation effects in a continuum of temperatures, T$_{el}$, in the static part of the calculation. The effect of T$_{el}$ on the vibrational properties is more of a second-order effect, even though here we see it is crucial to stabilizing the B2-IS state. Despite this non-negligible effect, once phonons stabilize, the impact of T$_{el}$ on the vibrational free energy should be less significant than on the static energy. Here we perform the static calculation in a Tel continuum. The electronic temperatures are sampled from 0 K to 7,000 K spaced by 1,000 K with temperature interpolations. With the vibrational entropy S$_{vib}$ from phonon dispersion, the Gibbs free energy of B1 and B2 phases can be computed using the quasi-harmonic approximation calculations. Obviously, T$_{el}$ when the system is in thermodynamic equilibrium. A common expression for the free energy in this case
\begin{equation}
F(V,T,T_{el})=F_{static}(V,T_{el})+F_{vib}(V,T,T_{el})-TS_{mag}, \label{2a}
\end{equation}

where

\begin{equation}
F_{static}(V,T_{el})=F_{Mermin}(V,T_{el}), \label{2b}
\end{equation}

is the total Mermin free energy at volume V. Here,

\begin{equation}
F_{Mermin}(V,T_{el})=E_{static}(V,T_{el})-T_{el}S_{el}(V,T_{el}), \label{2c}
\end{equation}

where E$_{static}$ (V,T$_{el}$) is the self-consistent energy with orbital occupancies

\begin{equation}
f_{\textbf{k}i}(V,T_{el})=\frac{1}{\exp{\frac{\hbar(E_{\textbf{k}i}-E_{F})}{k_BT_{el}}}+1}, \label{2d}
\end{equation}

with E$_{\textbf{k}i}$ being the one-electron energy of an orbital with wavenumber \textbf{k} and band index i, and E$_F$ being the Fermi energy. The electronic entropy is

\begin{equation}
S_{el}=-k_B\sum_{\textbf{k},i}[(1-f_{\textbf{k}i})\ln(1-f_{\textbf{k}i})+f_{\textbf{k}i}\ln{f_{\textbf{k}i}}]. \label{2e}
\end{equation}

The vibrational energy is

\begin{widetext}

\begin{equation}
\begin{aligned}
F_{vib}(V,T,T_{el})=\frac{1}{2}\sum_{\textbf{q},s}\hbar\omega_{\textbf{q},s}(V,T_{el}=0)+ 
k_BT\sum_{\textbf{q},s}\ln{\left\{1-\exp{[-\frac{\hbar\omega_{\textbf{q},s}(V,T_{el})}{k_BT}}]\right\}}, \label{2f}
\end{aligned}
\end{equation}

\end{widetext}

where $\omega_{(\textbf{q},s)}$ (V) is the vibrational frequency of noninteracting phonons with wavenumber \textbf{q} and polarization index s.

Another important contribution to free energy in fp is the magnetic entropy\cite{23}, here treated in the atomic impurity limit,

\begin{equation}
S_{mag}=k_Bx_{Fe}n\ln{[m(2S+1)]}, \label{2g}
\end{equation}

where k$_B$ is the Boltzmann constant, x$_{Fe}$ is iron concentration, \textit{n} is the fraction of B2-IS states, S is the iron total spin quantum number (S=0 for LS and S=1 for IS), and \textit{m} is the orbital configuration multiplicity. For an insulating system, it is easier to count this multiplicity, and \textit{m} would be three if the B2-IS state were insulating. However, in a calculation with the Mermin functional, which includes electronic entropy, with nearly degenerate t$_{2g}$ states at the Fermi level, it is more appropriate to define \textit{m}=1 for the B2-IS and LS states.

Next, we addressed the issue of configurational entropy. fp is a solid solution of FeO and MgO. According to the Gibbs phase rule, this leads to a B1-B2 coexistence region \cite{41}. Here we use the quasi-ideal solid solution model to determine the coexistence region and phase boundary. This method, detailed in the Method section, only needs the free energy information for MgO and (Fe$_{0.125}$Mg$_{0.875}$)O, and does not require the free energy data on the Fe-rich side (Valencia-Cardona et al., 2021, under review). We note that this method improves on the ideal solid-solution model to compute the phase boundary for Mg-rich concentrations. As shown in the B1 phase, in Fe-rich concentrations, complex Fe-Fe interactions may lead to non-uniform iron distribution and symmetry reduction \cite{45}.

\begin{figure}[t]
\includegraphics[width=0.48\textwidth]{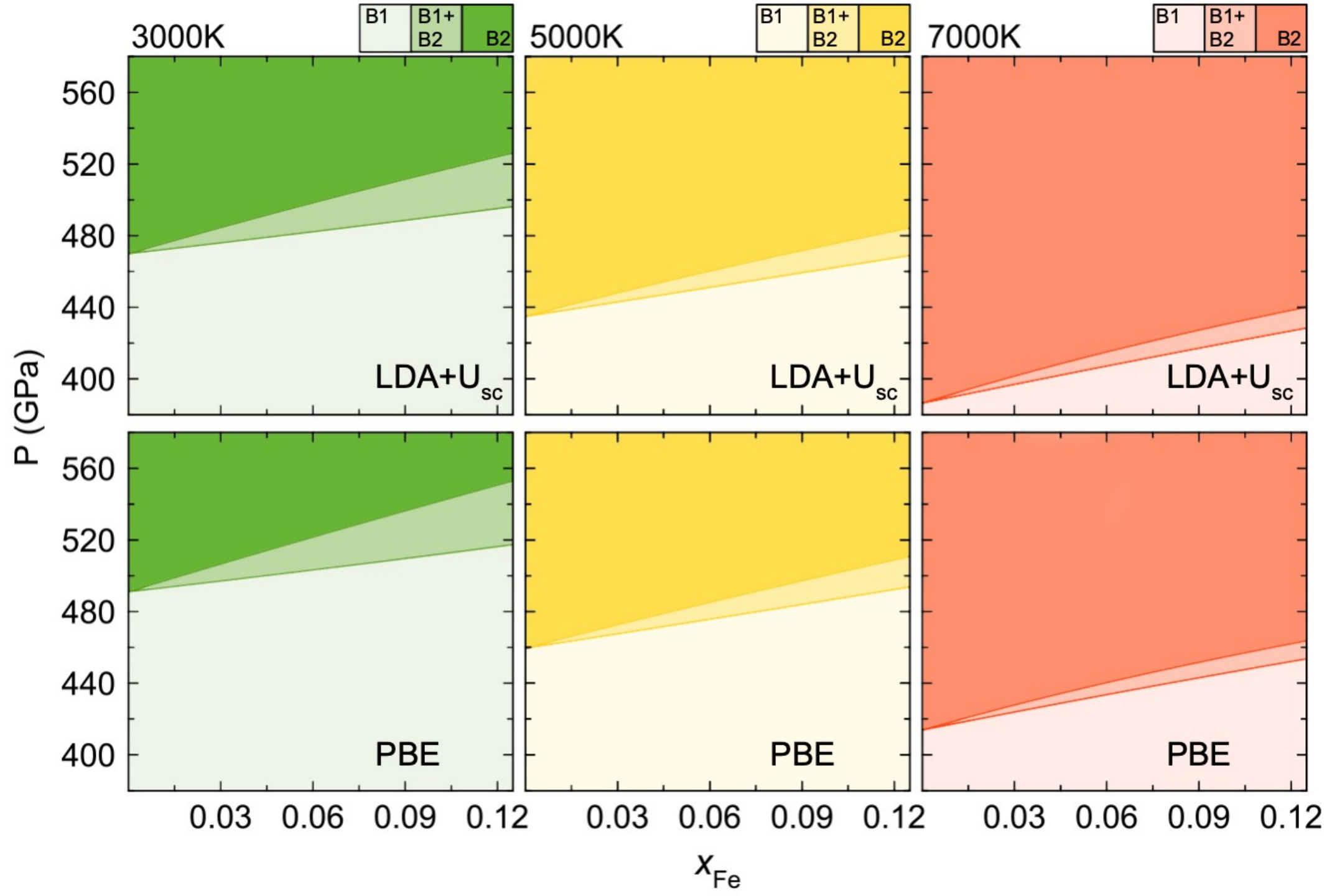}
\caption{\label{fig:fig6} Pressure-concentration phase diagram based on the quasi-ideal solid solution model.}
\end{figure}

Fig.~\ref{fig:fig6} shows the pressure-concentration phase diagram for the B1 $\rightarrow$ B2 transition at various temperatures. With increasing temperature, the phase boundary shifts to lower pressures. This is because the electronic entropy contributes more significantly to the stabilization of the B2 phase. Moreover, as shown in Fig.~\ref{fig:fig5}, the B2 phase has smaller vibrational frequencies than the B1 phase due to larger bond lengths, resulting in larger vibrational entropy for this phase. Therefore, these entropic effects stabilize the B2 phase at higher temperatures. The B1-B2 coexistence region widens with increasing iron concentration while it narrows down at higher temperatures. Results by LDA+U$_{SC}$ and PBE are qualitatively similar, except that the phase boundary by PBE is systematically shifted to higher pressures.

\begin{figure}[t]
\includegraphics[width=0.5\textwidth]{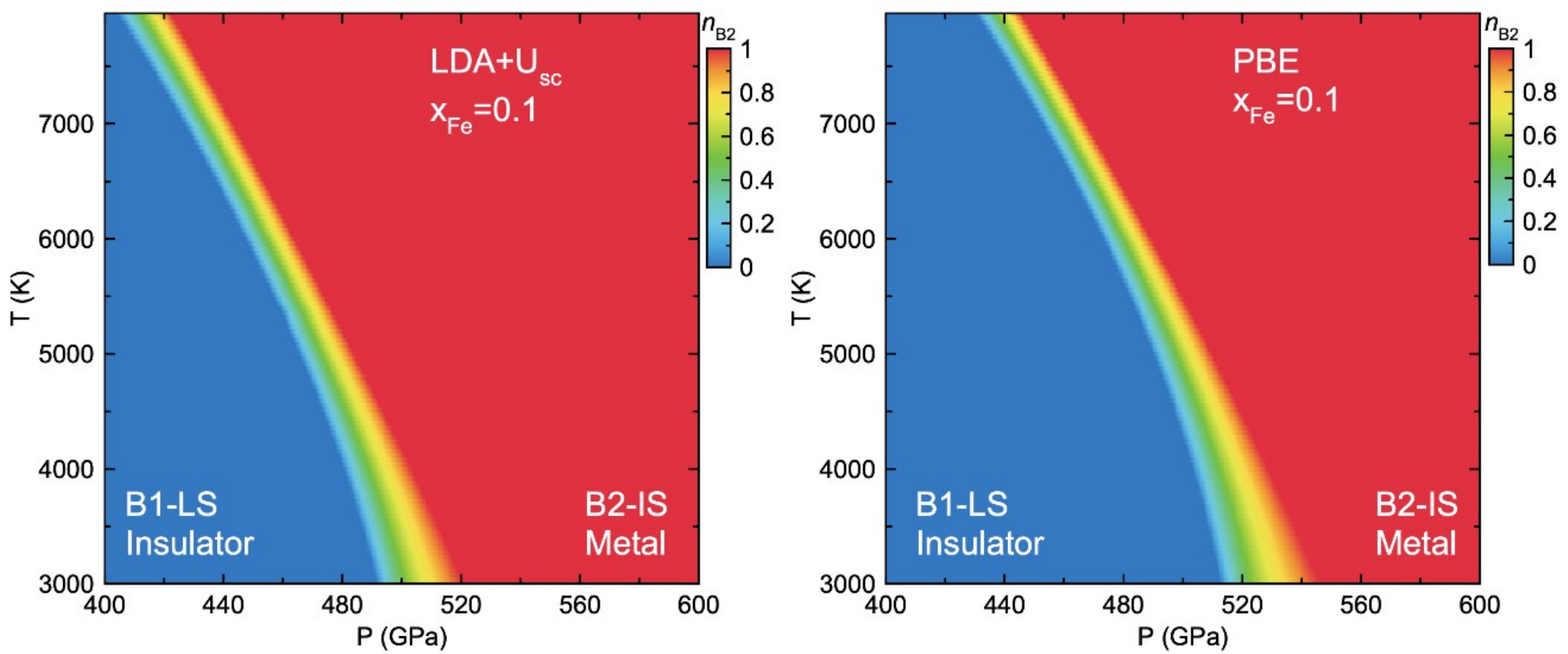}
\caption{\label{fig:fig7} Phase boundary between B1-LS and B2-IS with x${_Fe}$= 0.1. The color bar shows $n_{B2}$, the molar fraction of the B2-IS phase; (a) LDA+U$_{SC}$ results; (b) PBE results. The B2-LS state plays an insignificant role in this phase transition.}
\end{figure}

The P-T phase diagram for x$_{Fe}$= 0.1 is computed using the quasi-ideal solid-solution model and \myeqref{1c1} with both LDA+U$_{SC}$ and PBE functionals. Because \myeqref{2g} gives the maximum magnetic entropy in the metallic state, we also compute the phase diagram without this term, which provides the lower limit for S$_{mag}$. The phase boundaries with and without S$_{mag}$ are shown in Fig.~\ref{fig:figS7}. We find that S$_{mag}$ does not significantly affect the phase boundary but changes the width of the coexistence region. Therefore, we provide the phase boundary in Fig.~\ref{fig:fig7} using the average S$_{mag}$. The phase boundary shows a negative Clapeyron slope because entropic effects stabilize the B2 phase at high temperatures. This is consistent with the previous observations of large negative B1-B2 Clapeyron slope in both end members, FeO \cite{46} and MgO\cite{47}. Similar to static calculations, the high-temperature PBE calculations lead to a higher transition pressure of $\sim$ 20 GPa compared to LDA+U$_{SC}$ calculations.

Upon further compression, we observe the B2-IS $\rightarrow$ B2-LS state crossover. This spin-state change in the metallic system can be addressed by inspecting the evolution of the average magnetic moment per iron. With a small electronic temperature of T$_{el}$=1,000K, we find the B2-IS state is dynamically stable up to 3.0 TPa. The pressure-dependent magnetization displayed in Fig.~\ref{fig:fig8} shows no B2-IS $\rightarrow$ B2-LS transition in static calculations for T$_{el}$ = 1,000K. Interestingly, with rising T$_{el}$, the average magnetic moment reduces more significantly with increasing pressure. At T$_{el}$ = 5,000 K, the magnetic moment of fp12 in the B2 phase vanishes at P=2.75 TPa in LDA+U$_{SC}$ calculations. The PBE calculation also identifies a similar IS $\rightarrow$ LS transition for T$_{el}$ = 5,000 K at P = 1.63 TPa. Therefore, this spin-state change is much more sensitive to the functional than the B1 $\rightarrow$ B2 structural transition. Moreover, comparing different conditions in Fig.~\ref{fig:fig8}, one can see that increasing T$_{el}$ and x$_{Fe}$ lower the B2-IS $\rightarrow$ B2-LS transition pressure.

\section{IV. Discussion}
These calculations in B1 and B2 phases suggest several states are possible for Mg$_{1-x}$Fe$_x$O with  x$_{Fe}$ $\leq$ 0.125. The ultra-high pressure in super-Earths can lead to a complex, layered mantle structure. In the outer layers, one would expect B1-type fp. With increasing depth, fp first undergoes the HS-LS transition similar to that on Earth. Then, fp transforms from the B1 to the B2 structure. The B1-B2 coexistence region spreads over a finite pressure range, e.g., $\sim$ $\Delta$25 GPa at 3,000 K. Along with the structural transition, the re-emergence of a magnetic state with a simultaneous insulator-to-metal transition can have a strong impact on thermal and electrical conductivities, which will be important to model heat transport and convection in a multi-layered super Earth mantle. Coupling between the electromagnetic field produced in the core and the high magnetic induction of the metallic mantle could influence the temporal evolution of the magnetic field lines crossing the mantle. The metallization of the mantle could also affect super Earth's nutation \cite{48,49}. 

\textbf{
\begin{figure}[t]
\includegraphics[width=0.5\textwidth]{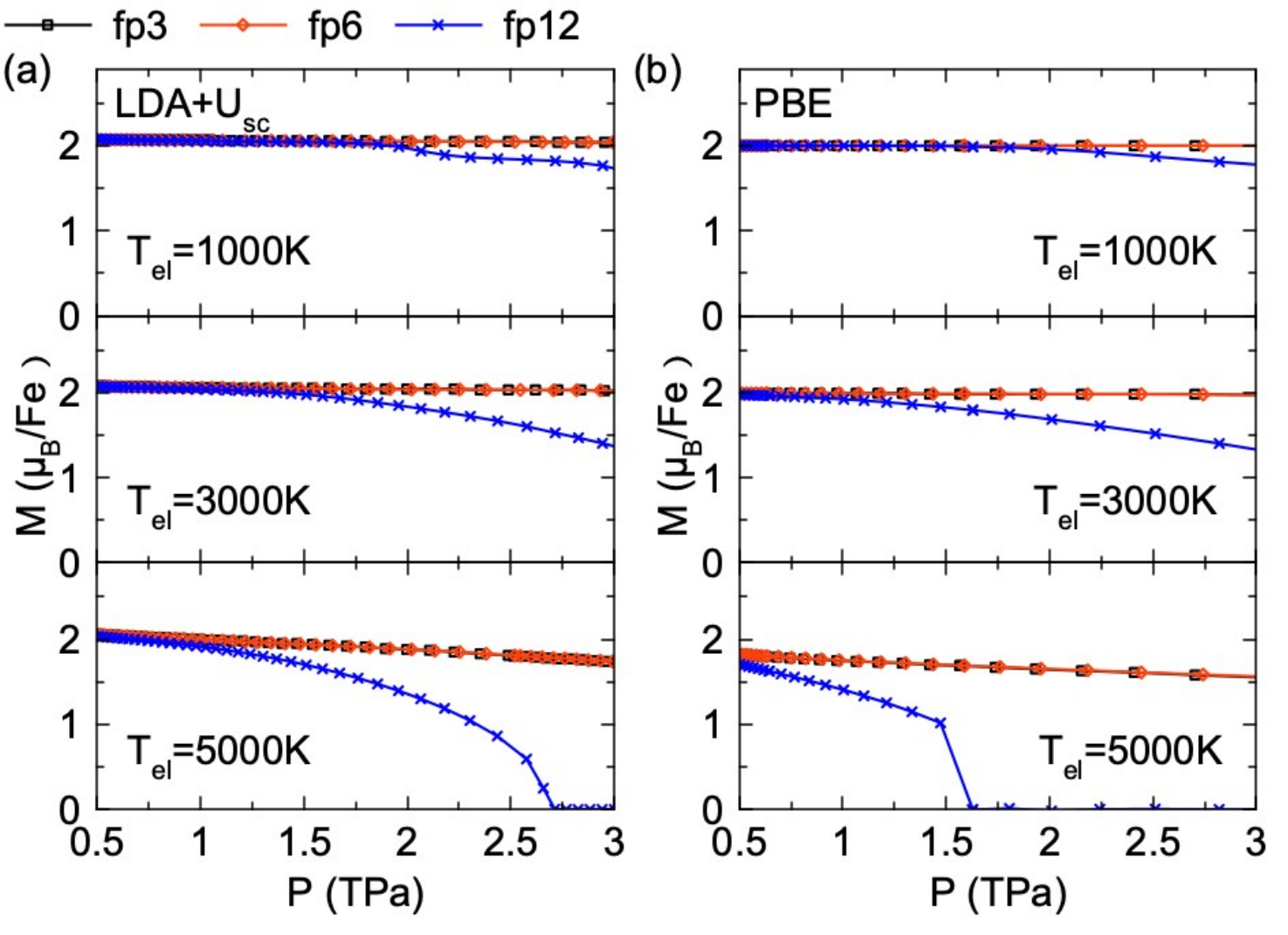}
\caption{\label{fig:fig8} Magnetic moment of iron vs. pressure in fp3, fp6, and fp12 computed with (a) LDA+U$_{SC}$ and (b) PBE. The temperatures for thermal electronic excitations, T$_{el}$, are indicated in each panel.}
\end{figure}
}

The current study focuses mainly on the solid solution with x$_{Fe}$ $\leq$ 0.125, which is relevant for fp. Because FeO's B1 $\rightarrow$ B2 transition pressure is lower than MgO’s but only exists above 4,000 K \cite{46}, the current trend of increasing transition pressure with increasing iron concentration should reverse at higher iron content. Iron-iron interactions may also cause the dissociation into a non-uniform solid solution, i.e., an iron-rich and an iron-poor phase separation in x$_{Fe}$ = 0.5 \cite{50}. Therefore, the phase transition at relevant higher iron concentrations needs a more robust thermodynamic treatment and is left for a future study.

While this paper was in preparation, a preprint \cite{51} on the same topic became available. That study focuses on the B1 $\rightarrow$ B2 transition in fp with higher iron concentrations, e.g., x$_{Fe}$ = 0.125 and 0.25. They used the LDA+U$_{SC}$ functional as well. First, their static transition pressure decreases with increasing x$_{Fe}$, which we confirm. This is in line with the lower transition pressure of FeO compared to that in MgO recently found by \cite{10}. The increase in transition pressure we observed for low x$_{Fe}$ values, suggests that fp becomes a non-ideal solid solution beyond x$_{Fe}$=0.125. Second, their fp12 B2-IS phase displays a rhombohedral distortion producing a semiconducting state with a small gap of 0.3 eV at T$_{el}$ = 0 K. This result is qualitatively similar to ours on fp3 before considering the effect of thermal electronic excitations. Therefore, their rhombohedra phase should also become metallic at high temperatures and possibly also turn into cubic. The LDA+U$_{SC}$ calculation on the metallic state requires electronic entropic contributions to the internal energy using the Mermin functional \cite{22}. Considering that thermal electronic excitations can stabilize phonons in the B2-IS state of the solid solution, our calculations with the cubic cell should be more appropriate to describe the B1-B2 high-temperature phase diagram.

The present calculation has explored the important effects of strong electronic correlation, thermal electronic excitations, structural distortions, and insulator to metal transition on the phase stability of fp under pressure. However, given the complexity of the problem, it is likely that electron-phonon and phonon-phonon (anharmonicity) interactions still play an important role in the behavior of the MgO-FeO solid solution under pressure, more so at low temperatures. Experiments on fp with similar iron concentrations will be essential to help to calibrate these challenging calculations. 

\section{V. Conclusions}
In conclusion, \textit{ab initio} calculations reveal a B1 to B2 transition in fp accompanied by an insulator to metal transition and the re-emergence of a local magnetic moment in iron corresponding to S=1, the intermediate spin (IS) state. The formation of the IS state originates in the change in coordination from octahedral to cubic and the change in the energy level structure of the 3\textit{d} electrons in Fe. Based on LDA+U$_{SC}$ calculations, the B1-LS $\rightarrow$ B2-IS static transition pressures, Pt, varies from  526 GPa for  x$_{Fe}$=0.03125 to 576 GPa for  x$_{Fe}$ = 0.125. PBE calculations provide a similar trend while the transition pressure is systematically higher by $\sim$ 20-30 GPa. Phonon calculations confirm the dynamic stability of the B2-IS state in the currently studied pressure range. By including vibrational and electronic entropy contributions in the quasi-ideal solid solution model, we determine the B1-B2 phase boundary and the coexistence region up to 7,000 K. The two-phase region widens with increasing iron concentration. The P-T phase diagram shows a negative Clapeyron slope. At ultrahigh pressure, the stability of the B2-IS state highly depends on the electronic temperature and iron concentrations. With increasing electronic temperature or iron concentration, the Fe magnetic moment tends to decrease to zero, the LS state. An insulator to metal transition accompanies this B1-B2 phase change, and the iron magnetic moment re-emerges, which should have significant consequences for modeling super-Earth’s mantle structure and dynamics. Further studies with a more robust thermodynamic model are needed to model this transition in fp with higher iron concentrations x$_{Fe}$. For x$_{Fe}$ $\geq$ 0.5, the B8 phase might also have to be considered.

\section{acknowledgments}
This work is supported by the by National Science Foundation awards EAR-1918126. R.M.W. and Y.S. also acknowledge partial support from the Department of Energy, Theoretical Chemistry Program through grant DE-SC0019759. Computational resources were provided by the Extreme Science and Engineering Discovery Environment (XSEDE) funded by the National Science Foundation through award ACI-1548562. The authors also acknowledge the Texas Advanced Computing Center (TACC) at The University of Texas at Austin for providing high-performance computer resources that have contributed to the research results reported within this paper.

\twocolumngrid

\bibliographystyle{apsrev4-1}


\pagebreak
\widetext
\begin{center}
\textbf{\large Supplementary Information for ``Intermediate spin state and the B1-B2 transition in ferropericlase at tera-Pascal pressures''}
\end{center}

\setcounter{equation}{0}
\setcounter{figure}{0}
\setcounter{table}{0}
\renewcommand{\thefigure}{S\arabic{figure}}
\renewcommand{\thetable}{S\arabic{table}}

\begin{center}
\textbf{Content}\\
\end{center}
Supplementary Text Detailed explanation for quasi-ideal solid solution model
Figure S1. Projected density of states and phonon dispersions for B2-IS with x$_{Fe}$= 0.03125 from PBE calculations at $\sim$ 510 GPa. \\
Figure S2. Projected density of states and phonon dispersions for B2-IS with x$_{Fe}$= 0.125 from LDA+U$_{SC}$ calculations at $\sim$ 525 GPa. \\
Figure S3. Phonon dispersions for B2 x$_{Fe}$= 0.03125 with different electronic temperature at $\sim$ 510 GPa. \\
Figure S4. Phonon dispersions for B2 x$_{Fe}$= 0.125 with different electronic temperature at $\sim$ 525 GPa. \\
Figure S5. Atomic structure of supercells for B1 and B2 x$_{Fe}$= 0.0625 and 0.125. \\
Figure S6. Projected density of states for ferrous iron B2-LS at ~1TPa and B1-LS states at $\sim$ 450 GPa.\\
Figure S7. Phase boundary between B1-LS and B2-IS with x$_{Fe}$= 0.1. \\
Figure S8. Phase boundary between B1-LS and B2-IS with different electronic temperatures using PBE calculations. \\
Table S1. Equation of state parameters from LDA+U$_{SC}$ calculations.\\
Table S2. Equation of states parameters from PBE calculations.\\

\textbf{Supplementary Text: Detailed explanation for quasi-ideal solid solution model}\\
In principle, there is more than one level of solid solution modeling for fp. First, there is the FeO-MgO solution modeling, (Mg$_{1-x}$Fe$_{x}$)O, with variable FeO concentration and fixed iron spin state. This modeling is done separately for each spin state. It produces free energies vs. x$_Fe$, P, and T for (Mg$_{1-x}$Fe$_{x}$)O in a single spin state. Second, for a fixed x$_Fe$, there is the spin-crossover modeling, from intermediate spin (IS) to low spin (LS), where (Mg$_{1-x}$Fe$_{x}^{IS}$)O and (Mg$_{1-x}$Fe$_{x}^{LS}$)O are the end-member phases. For low iron concentrations, there is negligible Fe-Fe interaction (elastic or exchange), and the “ideal” solid solution (ISS) formalism can be applied to the end-member phases (Mg$_{1-x}$Fe$_{x}^{IS}$)O and (Mg$_{1-x}$Fe$_{x}^{LS}$)O.

In the context of a solid solution between MgO and FeO in a single spin state, the ideal solid solution model uses only the free energy of the end-member phases MgO and FeO and adds an ideal-mixing configuration entropic contribution. The interaction between MgO and FeO is completely disregarded. \\
\begin{equation}
\begin{aligned}
G = (1-x_{Fe})G_{MgO}  + x_{Fe} G_{FeO} + G_{mix},
\end{aligned}
\end{equation}

where $G_{mix}  = k_B T[x_{Fe} lnx_{Fe}  + (1-x_{Fe})ln(1-x_{Fe})]$ and G = G(P,T,x$_{Fe}$) for short. 

The interaction between MgO and FeO is naturally included in supercell calculations such as the one we have carried out. Therefore, our thermodynamic analysis goes beyond the ISS modeling. 

Our model uses MgO and (Mg$_{0.875}$Fe$_0.125$)O as end-member phases. The free energy of the latter is computed using only one configuration, which is adequate for systems with small x$_{Fe}$. The (Mg$_{0.875}$Fe$_0.125$)O free energy includes the “ideal” entropic contribution, G$_{mix}$, for x$_{Fe}$= 0.125. Using these two end-members, we model a solid solution of (Mg$_{1-x}$Fe$_{x}$)O, with x$_{Fe}$ $<$ 0.125, using the ISS formalism

\begin{equation}
\begin{aligned}
G = (1-x_{B})G_{MgO}  + x_{B} G_{(Mg_{0.875}Fe_{0.125})O}  + G_{mix}  - x_B G_{mix}^{fp12},
\end{aligned}
\end{equation}

where $G_{mix}   = k_B T[x_B lnx_B  + (1-x_B )ln(1-x_B )]$, $G_{mix}^{fp12}=k_B T[0.125ln0.125+0.875ln0.875]$. The actual iron concentration in this system is $x_{Fe}=x_B*0.125$.

This modified ISS modeling goes beyond the traditional ISS modeling by including an interaction between the standard end-member phases, MgO and FeO. We refer to this formalism as the quasi-ideal solid solution model (QISS). 

The spin-crossover modeling follows after the computations of the free energy of (Mg$_{1-x}$Fe$_{x}^{IS}$)O and (Mg$_{1-x}$Fe$_{x}^{LS}$)O.\\

\begin{equation}
\begin{aligned}
G(x,n) = (1-n)G_{(Mg_{1-x}Fe_{x}^{IS})O} + nG_{(Mg_{1-x}Fe_{x}^{LS})O} + G_{mix}
\end{aligned}
\end{equation}

where “n” is the fraction of iron in the LS state and $G_{mix}  = k_B T[nlnn + (1-n)ln(1-n)]$.

In summary, the QISS model mitigates the lack of interaction ISS modeling between MgO and FeO on the low x$_{Fe}$ side. The IS to LS spin-crossover phenomenon in the B2 phase happens at pressures much higher than that of the B1-B2 transition, it does not interfere with the latter, and can be computed as an ISS where the end-members are treated as two QISSs.

\newpage

\begin{figure}[hp]
\includegraphics[width=0.8\textwidth]{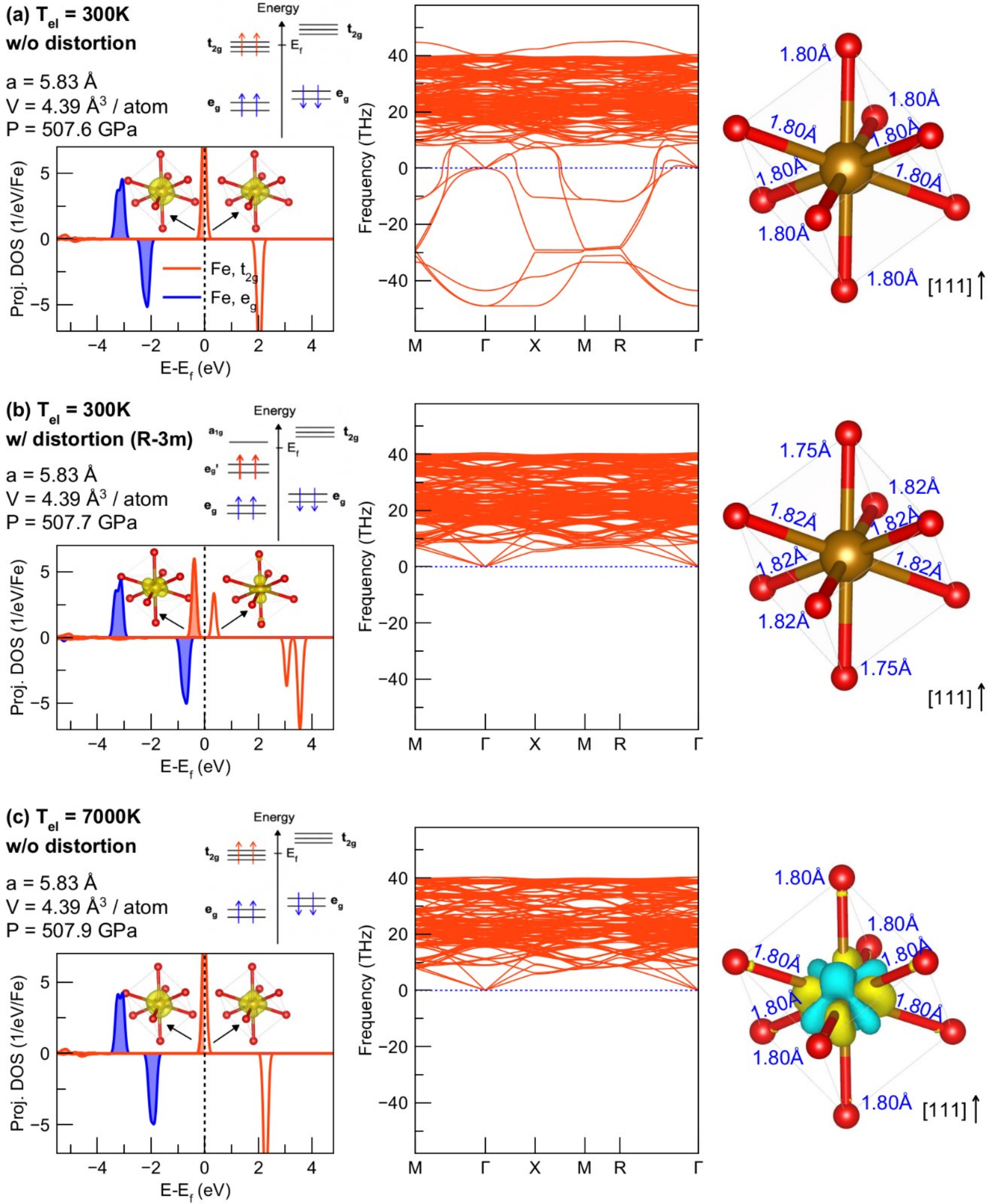}
\caption{\label{fig:figS1} Left panel: Projected density of states, charge density, and schematics of electron occupation of Fe$^{2+}$ in the B2-IS state with x$_{Fe}$= 0.03125 obtained with PBE. Middle: Phonon dispersion. Right panel: Fe-O bond length and Charge difference between (a) and (c). $(\Delta \rho=\rho(7000 K)-\rho(300 K)$. Charge changes from blue to yellow regions.)}
\end{figure}

\begin{figure}[hp]
\includegraphics[width=0.8\textwidth]{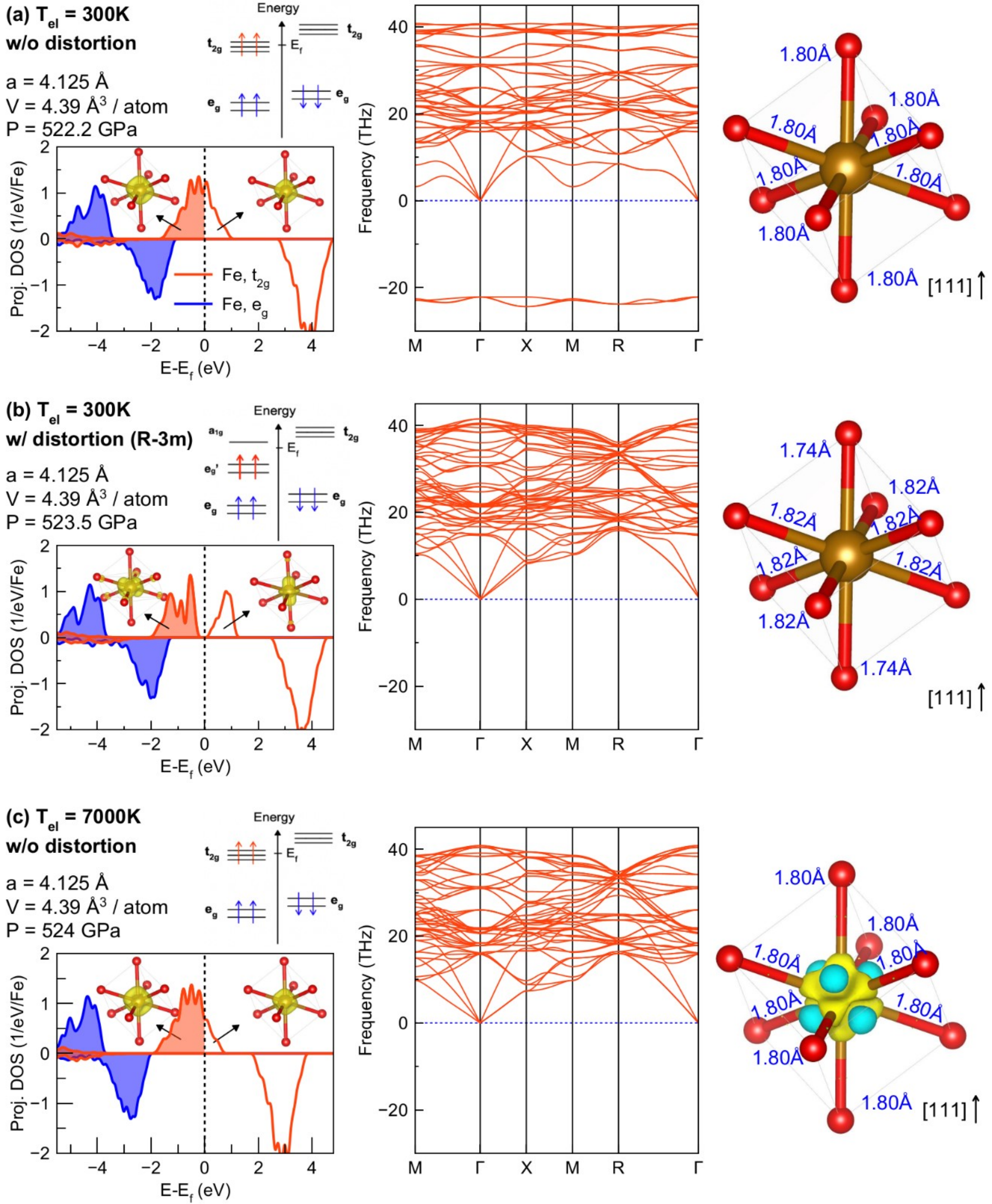}
\caption{\label{fig:figS2} Left panel: Projected density of states, charge density, and schematics of electron occupation of Fe$^{2+}$ in the B2-IS state with x$_{Fe}$= 0.125 obtained with LDA+U$_{SC}$. Middle: Phonon dispersion. Right panel: Fe-O bond length and Charge difference between (a) and (c). $(\Delta \rho=\rho(7000 K)-\rho(300 K)$. Charge changes from blue to yellow regions.)}
\end{figure}

\begin{figure}[hp]
\includegraphics[width=0.8\textwidth]{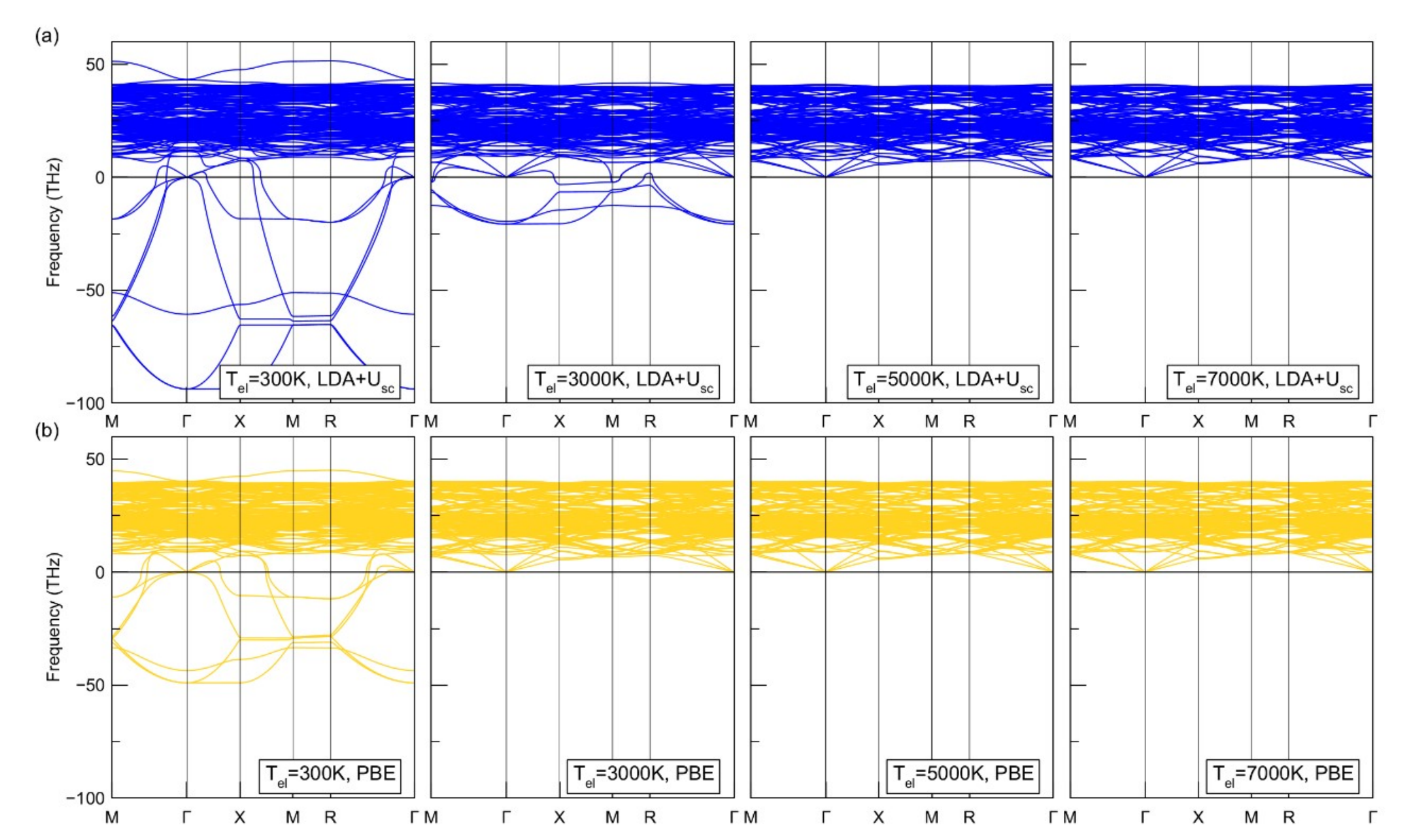}
\caption{\label{fig:figS3} Phonon dispersions at different electronic temperatures for the B2-IS state with x$_{Fe}$=0.03125 at $\sim$ 510 GPa computed using (a) LDA+U$_{SC}$ and (b) PBE functionals. With increasing electronic temperature, phonons are stabilized with LDA+U$_{SC}$ and PBE functionals. For T$_{el}$ \textgreater \, 5,000 K, LDA+U$_{SC}$ and PBE calculations provide similar results.}
\end{figure}

\begin{figure}[hp]
\includegraphics[width=0.8\textwidth]{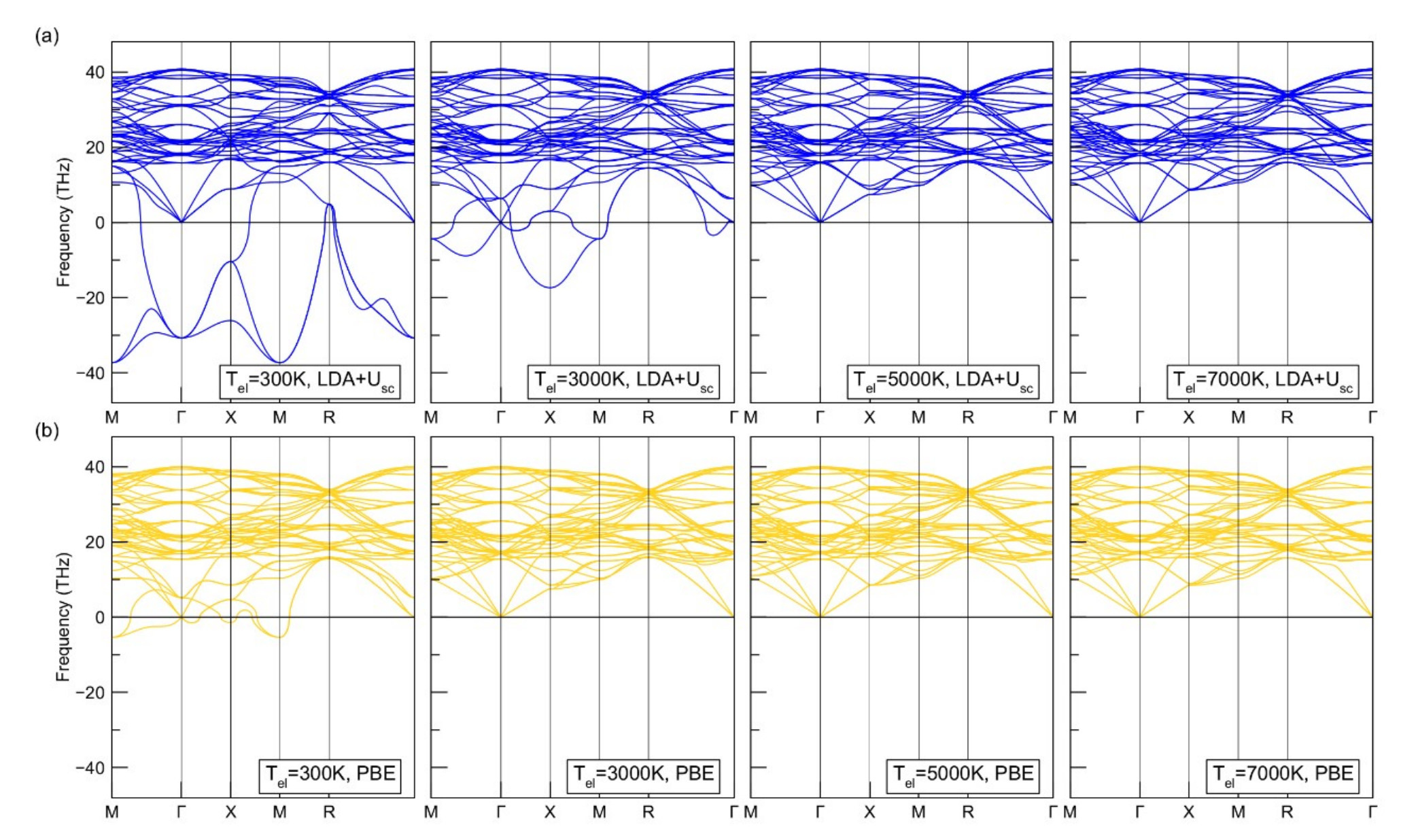}
\caption{\label{fig:figS4} Phonon dispersions at different electronic temperatures for the B2-IS state with x$_{Fe}$=0.125 at $\sim$ 525 GPa computed using (a) LDA+U$_{SC}$ and (b) PBE functionals. With increasing electronic temperature, phonons are stabilized with LDA+U$_{SC}$ and PBE functionals. For T$_{el}$ \textgreater \, 5,000 K, LDA+U$_{SC}$ and PBE calculations provide similar results.}
\end{figure}

\begin{figure}[hp]
\includegraphics[width=0.8\textwidth]{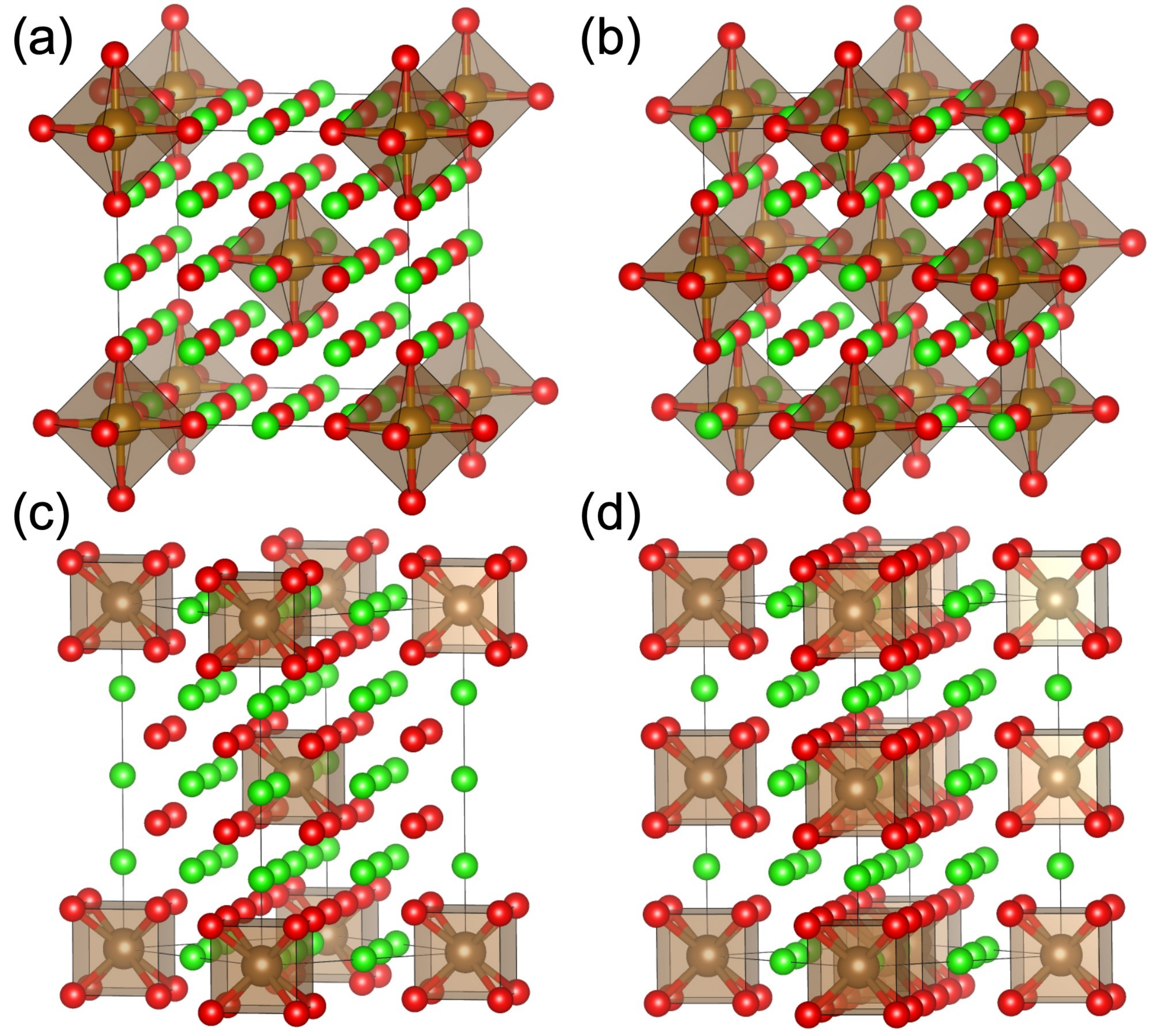}
\caption{\label{fig:figS5} Atomic structure of (a) B1 with x$_{Fe}$= 0.0625; (b) B1 with x$_{Fe}$ = 0.125; (c) B2 with x$_{Fe}$ = 0.0625; (d) B2 with x$_{Fe}$ = 0.125.}
\end{figure}

\begin{figure}[hp]
\includegraphics[width=0.8\textwidth]{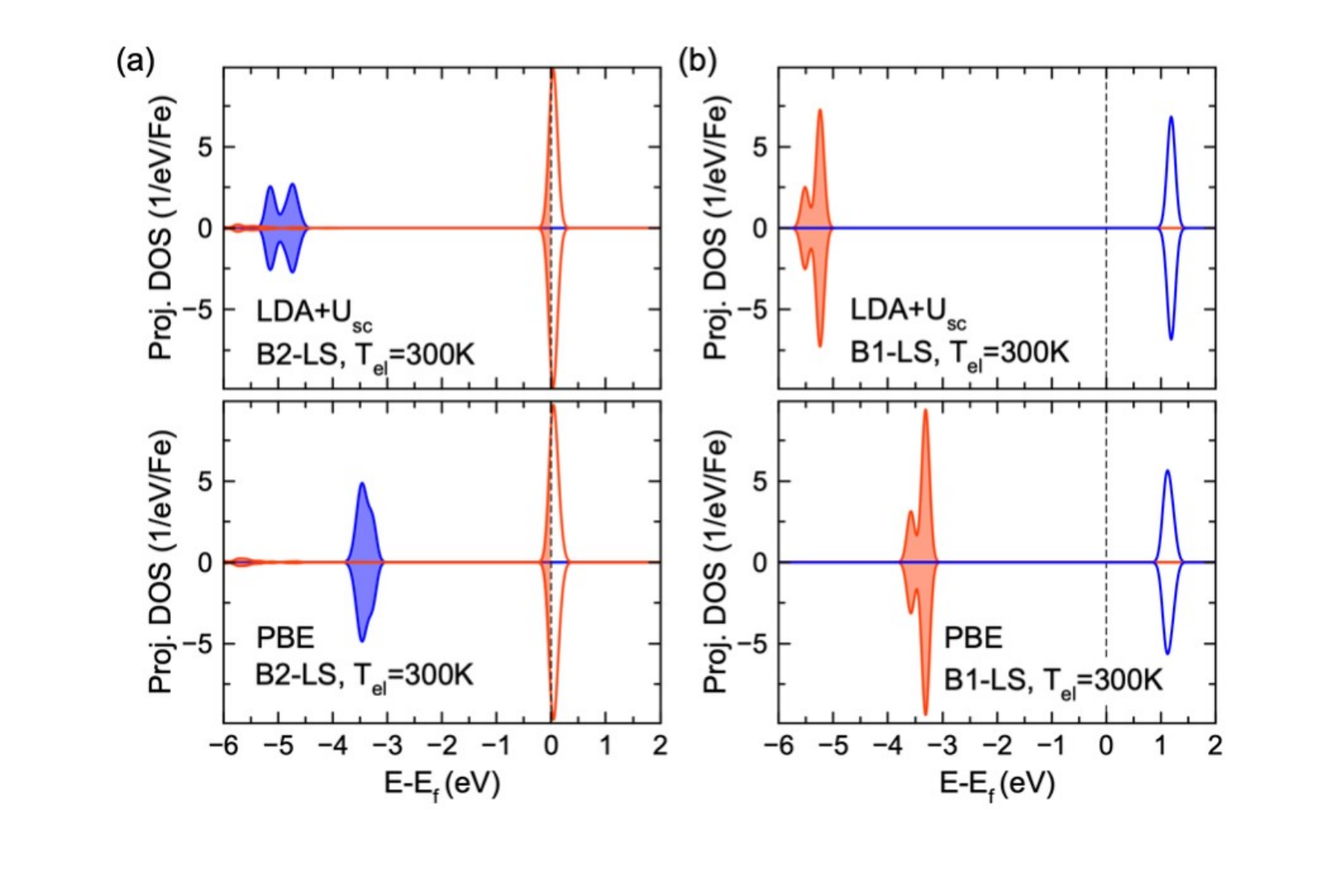}
\caption{\label{fig:figS6} Projected density of states for ferrous iron with x$_{Fe}$= 0.03125 in (a) B2-LS at 1TPa and (b) B1-LS at 450 GPa.}
\end{figure}

\begin{figure}[hp]
\includegraphics[width=0.8\textwidth]{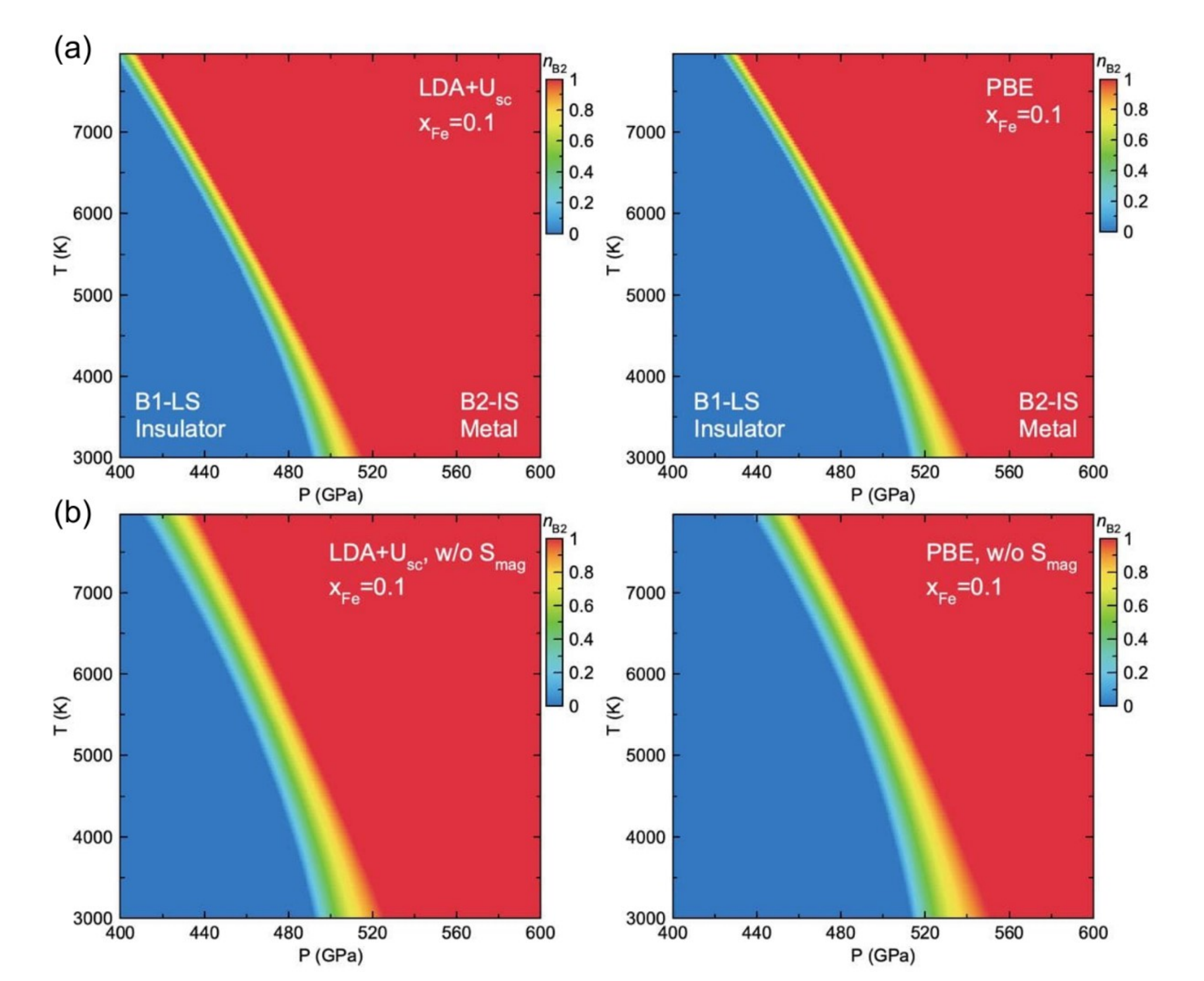}
\caption{\label{fig:figS7} Phase boundary between B1-LS and B2-IS for ferropericlase with x$_{Fe}$= 0.1 using both LDA+U$_{SC}$ and PBE functionals. The color bar shows $n_{B2}$, the molar fraction of the B2-IS phase. (a) Includes the effect of magnetic entropy; (b) disregards the effect of magnetic entropy.}
\end{figure}

\begin{figure}[hp]
\includegraphics[width=0.8\textwidth]{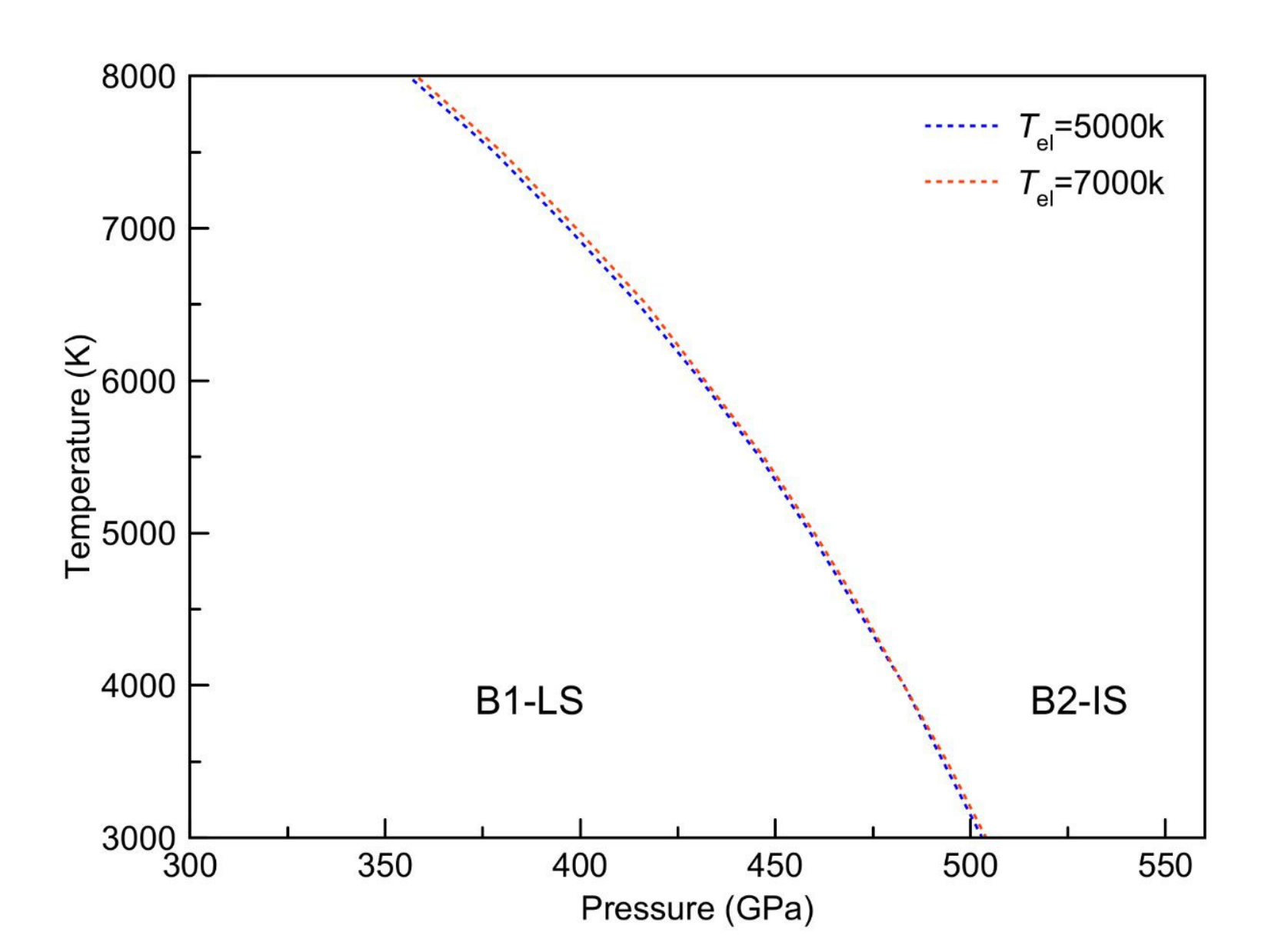}
\caption{\label{fig:figS8} Phase boundary between B1-LS and B2-IS with different electronic temperatures using PBE calculations.}
\end{figure}

\clearpage

\linespread{1.5}
\begin{table}[h]
    \centering
    \caption{LDA+U$_{SC}$ Birch-Murnaghan equation of state parameters for fp in different states. V$_0$ is the equilibrium volume, K$_0$ is bulk modulus while K$_0^{'}$ is its pressure derivative.}

\begin{tabular}{p{6cm}<{\centering}p{2.5cm}<{\centering}p{2.5cm}<{\centering}p{2cm}<{\centering}}
\hline
\hline \textbf{T=0K(static),300\textless P\textless650 GPa}  &  \textbf{V$_0$(\r{A}$^3$/atom)}  &  \textbf{K$_0$(GPa)}  &  \textbf{K$_0^{'}$}\\
\hline fp3-B1-LS&8.91&208.1&3.93\\
\hline fp6-B1-LS&8.88&210.4&3.94\\
\hline fp12-B1-LS&8.83&215.3&3.95\\
\hline fp3-B2-LS&8.68&195.1&3.95\\
\hline fp6-B2-LS&8.65&196.3&3.96\\
\hline fp12-B2-LS&8.63&201.4&3.97\\
\hline
\hline \textbf{T=0K(qha),300\textless P\textless650 GPa}  &  \textbf{V$_0$(\r{A}$^3$/atom)}  &  \textbf{K$_0$(GPa)}  &  \textbf{K$_0^{'}$}\\
\hline fp3-B1-LS&9.04&199.6&3.93\\
\hline fp6-B1-LS&8.94&206.2&3.94\\
\hline fp12-B1-LS&8.85&213.4&3.96\\
\hline fp3-B2-LS&8.86&183.3&3.96\\
\hline fp6-B2-LS&8.81&185.2&3.98\\
\hline fp12-B2-LS&8.75&189.3&4.01\\
\hline
\hline
\end{tabular}

    \label{tab:my_label1}
\end{table}

\linespread{1.5}
\begin{table}[h]
    \centering
    \caption{PBE Birch-Murnaghan equation of state parameters for fp in different states. V$_0$ is the equilibrium volume, K$_0$ is bulk modulus while K$_0^{'}$ is its pressure derivative.}

\begin{tabular}{p{6cm}<{\centering}p{2.5cm}<{\centering}p{2.5cm}<{\centering}p{2cm}<{\centering}}
\hline
\hline \textbf{T=0K(static),300\textless P\textless650 GPa}  &  \textbf{V$_0$(\r{A}$^3$/atom)}  &  \textbf{K$_0$(GPa)}  &  \textbf{K$_0^{'}$}\\
\hline fp3-B1-LS&9.26&180.8&3.94\\
\hline fp6-B1-LS&9.23&182.2&3.95\\
\hline fp12-B1-LS&9.18&184.9&3.98\\
\hline fp3-B2-LS&9.12&165.6&3.92\\
\hline fp6-B2-LS&9.09&167.5&3.94\\
\hline fp12-B2-LS&9.06&170.8&3.96\\
\hline
\hline \textbf{T=0K(qha),300\textless P\textless650 GPa}  &  \textbf{V$_0$(\r{A}$^3$/atom)}  &  \textbf{K$_0$(GPa)}  &  \textbf{K$_0^{'}$}\\
\hline fp3-B1-LS&9.36&176.9&3.93\\
\hline fp6-B1-LS&9.28&179.7&3.95\\
\hline fp12-B1-LS&9.20&184.5&3.97\\
\hline fp3-B2-LS&9.33&154.9&3.93\\
\hline fp6-B2-LS&9.26&155.4&3.95\\
\hline fp12-B2-LS&9.21&159.8&3.99\\
\hline
\hline
\end{tabular}

    \label{tab:my_label1}
\end{table}

\end{document}